\documentclass[12pt,oneside,reqno]{amsart}

\theoremstyle{definition}

\theoremstyle{remark}

\numberwithin{equation}{section}

\usepackage{amssymb}
\usepackage{graphicx}
\usepackage{bm}

\begin{document}

\title{Magnetostatic Spin Waves and %
Magnetic-Wave Chaos in Ferromagnetic Films. \\ %
I. Theory of Magnetostatic Waves in Plates %
under Arbitrary Anisotropy and External Fields}

\author{Yu.E. Kuzovlev, Yu.V. Medvedev, and N.I. Mezin}
\address{Donetsk Institute for Physics and Technology of NASU,
ul.\,R.\,Luxemburg 72, Donetsk 83114, Ukraine}
\email{kuzovlev@fti.dn.ua}





\begin{abstract}
General phenomenological theory of magnetic spin waves in
ferromagnetic media is originally reformulated   and applied to
analysis of magnetostatic waves in films and plates with arbitrary
anisotropy under arbitrary external field. Exact expressions are
derived for propagator of linear waves between antennae (inductors)
and mutual impedances of antennae, and exact unified dispersion
equation is obtained which describes all types of magnetostatic wave
eigen-modes. Characteristic frequencies (spectra) and some other
important properties of main modes are analytically considered and
graphically illustrated. Besides, several aspects of non-linear
excitations and magnetic-wave chaos are discussed, including
two-dimensional ``non-linear Schr\"{o}dinger equations'' for
magnetostatic wave packets and envelope solitons.
\end{abstract}


\maketitle

\baselineskip 24 pt

\markboth{}{}



\section{Introduction}

This preprint contains some results of research %
performed in %
Department of kinetic properties of disordered %
and nonlinear systems of  DonPTI NASU, %
between  2001 and 2003, %
and devoted to excitation and %
synchronization of chaotic magnetostatic spin waves (MSW) %
in ferromagnetic films. %
In fact, this is a part of report written at that time. %
It reflects our theoretical MSW's considerations   %
which, in our opinion, may be useful for concerned readers. %
Next preprints will reflect our numeric simulations %
of MSW and magnetic-wave chaos in films and corresponding %
real experiments. %

It should be noticed that a portion %
of the following Sections 2 and 3 already was presented %
(in a modified form) by arXiv preprint %
[Yu.\,Kuzovlev, N.\,Mezin, and G.\,Yarosh,\, %
cond-mat/0405640 ], %
but we keep it here for better reader's accommodation. %


\section{Basic properties and %
vocabulary of magnetic waves}

2.1. MAGNETIC TORQUE EQUATION.

In the classical phenomenology of ferro- and ferri-magnetic solids [1-6],
their magnetization is characterized by 3-dimensional %
vector field, \,$\,M(r,t)\,$\,,
representing density of elementary magnetic moments. The important
property of this vector is that its length is constant: \,$\,\,|M(r,t)|=M_{s}\,\,$\,%
, with \,$\,M_{s}=const\,\,$\, termed saturation magnetization. %
Hence, it is natural %
to measure magnetization and magnetic fields in \,$\,M_{s}\,\,$\, units and
introduce the unit-length ``spin'' vector \,$\,S(r,t)\,\,$\, by means of \,$\,%
M=M_{s}\,S\,$\,, \,$\,|S|=1\,$\,. Then natural time unit %
will be \,$\,\,\tau _{0}=(2\pi gM_{s})^{-1}\,\,$\,, %
where \,$\,\,g\,$\,  is gyromagnetic ratio (\,$\,g\approx 2.8\,\,$\, %
MHz/Oe\,) [1-3]. By this definition, \,$\,2\pi \tau _{0}\,$\, %
is period of %
precession of a separate spin at field equal to \,$\,M_{s}\,\,$\,. The
rotation of \,$\, S(r,t)\,\,$\, is subject to the Landau-Lifshitz torque equation
[1-6] which in these units reads
\begin{equation}
\frac{dS}{dt}=[F,S]+\gamma \{{F-S\langle S,F\rangle \}}\,\,,\,  F=-\frac{%
\delta E}{\delta S}\,\,,
\end{equation}
with \,$\,E\,\,$\, being magnetic energy functional %
and \,$\,\,F\,\,$\, full effective local magnetic %
field acting on spins (called also local thermodynamic force). %
Here and below  \,$\,\langle a,b\rangle \,$\,, \,$\,\,[a\,,b]\,\,$\,%
 and \,$\,\,a\otimes b\,\,$\, denote scalar product, vector product and tensor
product of two vectors  respectively %
(\,$\, c=\,a\otimes b\, $\, %
is matrix with elements \,$\,\,c_{\alpha \beta }=\, %
a_{\alpha }\,b_{\beta }\,$\, ). %
After transition to dimensionless time, \,$\,E\,\,$\, is factual %
energy divided by \,$\,M_{s}^{2}\,\,$\,, %
hence, its dimensionality is spatial volume,
while \,$\,F\,\,$\, becomes dimensionless. %
The term containing \,$\,\,\gamma \,$\, %
represents ``friction'', i.e. dissipative interaction between magnetization
and other microscopic degrees of freedom (thermostat). It can be written in
several equivalent forms  in view of identities
\[
F-S\langle S,F\rangle \,=\,[S,[F,S]] %
\,=\,(1-S\otimes S)\,F\,
\]
In principle, the Eq.1 is the only possible law of magnetic dynamics
compatible simultaneously with the constancy \,$\,|S|=1\,$\,, energy conservation
under spin rotation itself and total energy balance with
thermostat.\,  \,  In non-principally complicated models  the
friction may depend on spatial gradients %
of magnetization, \,$\,\nabla S(r,t)\,\,$\,.

2.2. FERROMAGNETIC ENERGY.

In any magnet, its energy functional \,$\, E\,\,$\, includes

(I) anisotropy energy

\begin{equation}
E_{a}\{S\}=\int A(S)\, dr\,
\end{equation}
which dictates locally preferred magnetization orientation (\,$\, dr\,\,$\, is
volume differential);

(II) exchange interaction energy, which ensures local spatial uniformity of \,$\,%
\,S(r)\,\,$\, pattern,

\begin{equation}
E_{e}\{S\}\,=\,\frac{1}{2}\, r_{0}^{2} %
\int \left\langle \nabla _{\alpha }\,S,\nabla
_{\alpha }\,S\right\rangle \,dr\,\,;
\end{equation}
it is invariant with respect to \,$\,\,S\,\,$\, rotation; in general, it may be
anisotropic with respect to gradient;

(III) dipole interaction energy which causes long-range non-uniformity
(demagnetization) of \,$\,\,S(r)\,\,$\, pattern and initializes formation of
magnetic domains:

\begin{equation}
E_{d}\{S\}\,=\,\frac{1}{2}\int %
\langle S,\widehat{G}\,S\rangle \,dr\,,
\end{equation}
with \,  \,$\,\widehat{G}\,\,$\, being the integral operator,

\begin{equation}
\widehat{G}\, S(r)\,\equiv \int %
G(r-r\prime )\,S(r\prime ) \,dr\prime \,,
\end{equation}
and \,$\,\,G(r)\,\,$\, the \,$\,3\times 3\,$\,-matrix function

\begin{equation}
G_{\alpha \beta }(r)\,=\, \frac{\delta _{\alpha \beta }} %
{|r|^{3}}-\frac{ %
3r_{\alpha }r_{\beta }}{|r|^{5}}\, =\, %
-\nabla _{\alpha }\nabla _{\beta }\frac{1}{ |r|}\,  ;
\end{equation}

(IV) interaction with an external magnetic field \,$\,H_{e}(r,t)\,$\,,

\[
E_{h}\,=\, -\int \langle H_{e}\,S\rangle \,dr
\]

In summary,
\[
E\,=\, E\{S,H_{e}\}\,=\, E_{a}+E_{e}+E_{d}+E_{h}+...
\]
The dots replace additional interactions  for instance, with distortions and
defects of atomic lattice. As a rule, their contribution to long magnetic
waves related phenomena (to be under our interest) can be either neglected
or reduced to renormalization of \,$\,M_{s}\,\,$\,, \,$\,\gamma \, \,$\,or other material
parameters [1,2,5]. From experiments it is known that especial surface
anisotropy may cause pinning of magnetization on the sample surface. This
effect can be described in terms of boundary conditions for the spin field \,$\,%
S(r,t)\,\,$ [8].

For example, in good quality yttrium-iron garnet (YIG) ferrites [4,5], \,$\,%
\,M_{s}\approx 140\,\,$\,Oe, correspondingly \,$\,\,\tau _{0}\approx 0.4\,\,$\,ns  the
exchange interaction radius \,$\,r_{0}\,$\,  \,  \,$\,\sim \,$\,  \,$\,%
5\cdot 10^{-6}\,\,$\,cm, characteristic anisotropy field is less than \,$\,%
\,M_{s}\, \,$\, (that is \,$\,|\partial A(S)/\partial S|\,\,$\, \,$\,\lesssim 1\,\,$\,), and the
dimensionless friction coefficient \,$\,\gamma \,$\, is rather small, \,$\,\gamma \sim
\,$\,  \,$\,10^{-4}\,\,$\,.

2.3. DIPOLE INTERACTION.

The dipole contribution (4) is nothing but approximate form of complete
magnetization interaction with self-induced %
magnetic field, \,$\,H_{S}(r,t)\,\,$\,. %
Usually, characteristic spatial scales (sample size, wavelength) and group
velocity of magnetic waves are much less than length and velocity,
respectively, of free-space electromagnetic wave with the same frequency.
Therefore, \,$\,H_{S}(r,t)\,$\, can be obtained as solution to quasi-static
Maxwell equations
\begin{equation}
\text{{\bf div}\,  }\,(H_{S}+4\pi S\,)\,=\,0\,,\,\,\, %
\text{{\bf curl}%
\,  }\,H_{S}\,=\,0\,,
\end{equation}
even in high-frequency (microwave) band. Besides  if there are no metal
surfaces in the vicinity of magnetic sample (waveguide walls or other), one
can suppose \,$\,H_{S}\rightarrow 0\,\,$\, at infinity and come to (4).
Oppositely, Eqs.7 should be solved under actual boundary conditions at
surroundings  and expression (4) be generalized as %
\,$\,\,E_{d}\,=\,-\int \langle H_{S}\,S\rangle \,dr/2\, \,$\,.

The principal peculiarity of dipole interaction is its specific long-range
character: with increasing distance \,$\,|r|\,$\, it decreases exactly as the volume
marked by this distance grows  \,$\,G(r)\,$\, \,$\,\propto r^{-3}\,$\,. Consequently, the
integral dipole force (5) is indifferent to absolute spatial scales of
magnetization pattern and depends on their ratios only. In particular, for
uniform magnetization it depends on a shape of magnetic sample but not on
its size.

2.4. STATIC MAGNETIZATION.

Let \,$\,S_{0}(r)\,$\, stands for a static (equilibrium, metastable or unstable)
magnetization configuration in a constant external field, \,$\,\,H_{e}=H_{0}(r)\,$\,
. According to Eq.1, the essence of static state is that everywhere spins
are oriented in parallel to thermodynamic force, \,$\,\,F\Vert S\,$\,, that is

\begin{equation}
-\frac{\delta E\{S_{0}\,H_{0}\}}{\delta S_{0}}\,=\,W_{0}(r)\,S_{0}\,,
\end{equation}
where scalar function \,$\,W_{0}(r)\,$\,(complete internal magnetic field) is
determined by the requirement \,$\,\,|S_{0}|=1\,$\,. Generally speaking, this
equation possesses many solutions (various domain structures) whose physical
realization depends on history of the system. But if \,$\,H_{0}\,\,$\, is
sufficiently strong then \,$\,\,$\,an unique solution must stay only [1], even in
spite of the dipole interaction, since the dipole force (5) %
is bounded,  \,$\,|\widehat{G}\,S|\leq 4\pi \,$\,, %
at any nonsingular magnetization
distribution.

2.5. MAGNETIC EXCITATIONS.

Consider perturbation of static state, \,$\,\,s\,=\,S-S_{0}\,\,$\,, for instance, caused
by additional time-varying external field, \,$\,\,h(r,t)\,\,$\,, %
when \,$\,H_{e}\,=\,H_{0}+h\,$\,. %
Under a perturbation, the energy divides
into two parts  \,$\,E=E_{\Vert }+E_{\bot }\, \,$\,, the ground
\[
E_{\Vert }\,=\,-\frac{1}{2}\int %
\{W_{0}+\langle H_{0}\,S_{0}\rangle \} \,dr \,-\, \int
\langle h,S_{0}\rangle \,dr\,,
\]
and excess energy \,$\,\,E_{\bot }\,$\, %
which vanishes at \,$\,\,s=0\,\,$\, and in its
turn consists of two parts

\begin{equation}
E_{\bot }\{s h\}\, =\frac{1}{2}\int %
W_{0}\,s^{2}\,dr \,+\, \widetilde{E}\{s h\}\, \,,
\end{equation}
\[
\widetilde{E}\{s h\}\,=\int \widetilde{A}(s)\,dr\,+\, %
E_{e}\{s\}\,+\, E_{d}\{s\}\, -\int %
\langle h,s\rangle\, dr\,
\]
The first contribution to (9) arises from interaction between  \,$\,s(r,t)\,\,$\,\
and static magnetization (lowering static magnetic order by magnetic
excitations  when \,$\,W_{0}(r)=\langle S_{0}\,F\{S_{0}\}\rangle \,$ %
$>0\,\,$, or, in opposite, maintaining it, when  \,$\, %
W_{0}(r)<0\,\,$\,). The second, %
\,$\,\widetilde{E}\{s h\}\,\,$\,, represents  \,$\,%
s(r,t)\,\,$\, interaction with itself. %
Its construction is quite similar to \,$\,%
E\{S,H\}\,$\, except that anisotropy term is modified:
\[
\widetilde{A}(s)=A(S_{0}+s)-A(S_{0})-\left\langle s A\prime
(S_{0})\right\rangle \,,\,  A\prime (S)\equiv %
\frac{\partial A(S)}{\partial S}\,
\]

This self-interaction induces the response operator \,$\,\,\widehat{L}\,\,$\,, as
defined by
\begin{equation}
\widehat{L}\,s\, \equiv \frac{\delta \widetilde{E}}{\delta s}\, %
+ h\,=A\prime
(S_{0}+s)-A\prime (S_{0})+\{\widehat{G}-r_{0}^{2}\nabla ^{2}\}\, s\,
\end{equation}
Evidently, in case of easy axis or easy plane anisotropy, when \,$\,A(S)\,$\, and
hence \,$\,\widetilde{A}(s)\,$\, are quadratic forms  \,$\,\widehat{L}\,\,$\, is purely
linear operator. In general, it is reasonable (and commonly used)
approximation if possible dependence of \,$\,\widehat{L}\,\,$\, on \,$\_{0}\,$\, is took
into account while its nonlinearity (dependence on \,$\,\,s\,\,$\, ) neglected,
thus replacing \,$\,\widehat{L}\,\,$\, by linear operator\,

\begin{equation}
\widehat{L}=\widehat{A}-r_{0}^{2}\nabla ^{2}+\widehat{G}\,,\, %
\,\text{where}\,\,  (\widehat{A}\,s)_{\alpha }\,=\, %
\frac{\partial ^{2}A(S_{0})}{\partial
S_{0\alpha }\partial S_{0\beta }}\, s_{\beta }\,.
\end{equation}

2.6. HAMILTONIAN FORMULATION.

Due to \,$\,|S_{0}+s|=1\,$\,, only two of three components of  the excitation  \,$\,%
s(r,t)\,\,$\, must be considered as independent variables  namely, the
components perpendicular to \,$\,S_{0}(r)\,$\,, since weak excitation is always
perpendicular to \,$\,S_{0}(r)\,$\,. Thus define the two-dimensional vector
\[
S_{\bot }=\widehat{\Pi }\,S\,\,,\, \widehat{\Pi }\equiv 1-S_{0}\otimes
S_{0}\,,
\]
where matrix  \,$\,\widehat{\Pi }=\widehat{\Pi }(r)\,\,$\, performs projection
onto the plane, \,$\,\Pi (r)\,\,$\,, %
perpendicular to \,$\,S_{0}(r)\,$\,. Evidently,
\begin{equation}
s\,=\,S_{\bot }+(S_{\Vert }-1)\,S_{0}\,\,,\, \,\,\, %
S_{\Vert }\,\equiv \langle S_{0},S\rangle \,\,,
\end{equation}
where scalar \,$\,\,S_{\Vert }\,\,$\, represents projection of  \,$\,\,S\,\,$\, onto the
static magnetization direction, and

\begin{equation}
S_{\Vert }^{2}+|S_{\bot }|^{2}=1\,\,, \,\,\,\, %
|s|^{2}\,=\,2(1-S_{\Vert })\,
\end{equation}
In terms of \,$\,S_{\bot }\,$\, the torque equation (1) transforms into

\begin{equation}
\frac{dS_{\bot }}{dt}\,=\, S_{\Vert }[F_{\bot }\,S_{0}]+\gamma (1{-}\,S_{\bot
}\otimes S_{\bot }{)}F_{\bot }\,\,,\,\,\,\, %
F_{\bot }\,=-\frac{\delta E_{\bot }\{s h\}}{\delta S_{\bot }}\,
\end{equation}
Here \,$\,\,E_{\bot }\{s h\}\, \,$\, should be expressed, with the help of (12)
and (13), as a functional of\, $\,\,S_{\bot }\,$\,, yielding

\begin{equation}
F_{\bot }\, =-\frac{S_{\bot }}{S_{\Vert }}\, %
\left( W_{0}-\left\langle S_{0}\,\frac{ %
\delta \widetilde{E}}{\delta s}\, %
\right\rangle \right) -S_{\Vert }\widehat{\Pi }\, %
\frac{\delta \widetilde{E}}{\delta s}\,\,\
\end{equation}

Notice that frictionless version of the Eq.14 (for \,$\,\gamma =0\,\,$\,) follows
from the variational principle

\begin{equation}
\delta \int \left( \int \left\langle S_{0}\, %
\left[ \frac{dS_{\bot }}{dt}\,%
,S_{\bot }\right] \right\rangle \, %
\frac{dr}{1+S_{\Vert }}\,+E_{\bot }\{s, h\}\right)\, dt\,=\,0
\end{equation}
It takes canonical Hamiltonian form after the change of variables

\begin{equation}
S_{\bot }\,=\,Q\sqrt{1-|Q|^{2}/4}\,\,,\, \,\, %
S_{\Vert }\,=\,1-|Q|^{2}/2\, \,,
\end{equation}
where vector \,  \,$\,Q\,$\, is also situated %
in the plane \,$\,\Pi (r)\,\,$\, [6-7]. %
Simultaneously such a choice of variables ensures unambiguous
parametrization of \,$\,\,S_{\Vert }\,\,$\, even in case of spin flipping when \,$\,%
\,S_{\Vert }\,\,$\, can become negative.\,  \,

2.7. DYNAMIC EQUATIONS FOR MAGNETIZATION.

It is convenient to introduce the rotation (precession) operator, \,$\,\widehat{R%
}\,\,$\,, and damp operator, \,$\,\,\widehat{\Gamma }\,\,$\,, by means of

\begin{equation}
\widehat{R}V\,=\,[S_{0}\,V]\,\,,\,\,\,\, %
\widehat{\Gamma }V\equiv \widehat{ \Gamma } %
(S_{\bot })V=\frac{\gamma }{S_{\Vert }}\widehat{\Pi }(1{-}\, %
S_{\bot }\otimes S_{\bot }{)}V{\,\,,}
\end{equation}
where\,$\,\,V\,\,$\,is arbitrary vector. Then the Eqs.14-15 takes more pragmatic
form:

\begin{equation}
\frac{dS_{\bot }}{dt}=\{\widehat{R}-\widehat{\Gamma }(S_{\bot
})\}\{(W_{0}+\langle S_{0}\,f\rangle )\,S_{\bot }-S_{\Vert }f\}\,\,,\,  \,\
f\equiv h-\widehat{L}\{S_{\bot }+(S_{\Vert }-1)\,S_{0}\}\,,
\end{equation}
to be supplemented by relations (11-13). Note that \,$\,\widehat{\Pi }^{2}\,$\, \,$\,=%
\widehat{\Pi }\,\,$\,,\,$\,\,\,\widehat{R}\widehat{\Pi }=\widehat{\Pi }\widehat{R}\,$\,
\,$\,=\widehat{R}\,\,$\,,\,$\, \widehat{R}^{2}=-\widehat{\Pi }\, \,$\,, and the latter
equality is equivalent to \,$\,\widehat{R}^{2}=\,$\, \,$\,-1\,$\, in planes \,$\,\Pi (r)\,\,$\,.

In generalized version of Eq.19, direct dipole interaction is replaced, in
analogy with (7), by self-induced field of the excitation, \,$\,h_{S}(r,t)\,$\,,
according to the rule \,$\, \,h_{S}\Leftrightarrow \,$\, \,$\,-\widehat{G}\, s \, \,$\,. In
such an approach

\begin{equation}
f=h+h_{S}-(\widehat{A}-r_{0}^{2}\nabla ^{2})\,s\,,
\end{equation}
and Eq.19 should be solved together with magnetostatic equations

\begin{equation}
\text{{\bf div}\,  }\,(h_{S}+4\pi s\,)=0\,,\,  \text{{\bf curl}%
\,  }\,h_{S}=0\,,
\end{equation}
under classical electromagnetic boundary conditions at the sample surface
and surrounding bodies.\,  \,

2.8. SMALL-AMPLITUDE DYNAMICS.

If a static state\,  \,  \,$\,S_{0}\,\,$\, is stable, i.e. represents
(global or local) energy minimum in the space of magnetization
configurations  then \,$\,E_{\bot }\{s h\}\,\,$\,, for small \,$\,s(r,t)\,\,$\,, is
positive quadratic form. Hence, we have rights to speak about
small-amplitude magnetic oscillations and waves. Linearization of Eq.19
results in

\begin{equation}
\frac{dS_{\bot }}{dt}=(\widehat{R}-\gamma \widehat{\Pi })(\widehat{W}\,S_{\bot
}-h)\,\,, \,  \widehat{W}\equiv W_{0}+\widehat{A}-r_{0}^{2}\nabla ^{2}+%
\widehat{G}\,,
\end{equation}
or, in the self-induced field representation,

\begin{equation}
\frac{dS_{\bot }}{dt}=(\widehat{R}-\gamma \widehat{\Pi })\{(W_{0}+\widehat{A}%
-r_{0}^{2}\nabla ^{2})\,S_{\bot }-h-h_{S}\}\,
\end{equation}

Let the perturbation and hence linear response vary harmonically: \,$\,h(r,t)=\,$\, \,$\,%
h(r)\exp (-i\omega t)\,$\,, and so on (\,$\,i\equiv \,$\, \,$\,\sqrt{-1})\,$\,. Then Eq.23
formally yields

\begin{equation}
S_{\bot }=\widehat{\chi }(h+h_{S})\,\,,\,  \widehat{\chi }=\widehat{\chi }%
(\omega )\equiv \{i\omega +(\widehat{R}-\gamma \widehat{\Pi })(W_{0}+%
\widehat{A}-r_{0}^{2}\nabla ^{2})\}^{-1}(\widehat{R}-\gamma \widehat{\Pi }%
)\,\,
\end{equation}
Substituting this expression to Eqs.21, we transform the problem to solving
partial differential equations

\begin{equation}
\text{{\bf div}\,  }\,\widehat{\mu }(h+h_{S}\,)=0\,,\, \widehat{%
\mu }(\omega )\equiv 1+4\pi \widehat{\chi }(\omega )\,\,,\, \text{{\bf %
curl}\,  }\,h_{S}=0\,
\end{equation}
Here \,$\,\widehat{\chi }\,$\, and \,$\,\widehat{\mu }\,$\, are play the role of linear
polarizability and susceptibility operators [2]. With respect to
polarization, \,$\,\widehat{\chi }\,$\, acts as \,$\,2\times 2\,$\,-matrix performing
projection onto planes \,$\,\Pi (r)\,$\,. It may be thought local spin precession
response to given local magnetic field, but, from the rigorous mathematical
point of view, presence of the Laplasian \,$\,\nabla ^{2}\,$\, makes it integral
(nonlocal) operator.\,  \,

2.9. LINEAR FREE WAVES.

Omitting in (22) friction and external pump, one can find eigenmodes and
eigen-frequencies of free magnetic waves (MW):

\begin{equation}
S_{\bot }\equiv Ve^{-i\omega t}\,,\,  -i\omega V=\widehat{R}\widehat{W}%
V\,\,
\end{equation}
It is sufficient to consider positive frequencies only, \,$\,\,\omega >0\,\,$\,,
since opposite sign means complex conjugating the same mode. Let different
modes be enumerated by a set of indexes \,$\,\,k\,\,$\,. Since operator \,$\,\widehat{W}%
\,\,$\, is positively defined, we can rewrite (26) as
\begin{equation}
\omega _{k}\widetilde{V}_{k}=i\widehat{W}^{1/2}\widehat{R}\widehat{W}^{1/2}%
\widetilde{V}_{k}\,,\, \,\widetilde{V}_{k}\equiv \widehat{W}^{1/2}V_{k}
\end{equation}
Operator in the left equation is Hermitian because \,$\,\,i\widehat{R}\, \,$\, is
Hermitian, hence, the solutions can be chosen mutually orthogonal and
normalized to \,$\,\,\int \langle \widetilde{V}_{m}^{\ast }\,\widetilde{V}%
_{k}\rangle dr\,$\, \,$\,=\omega _{k}\delta _{mk}\,\,$\, (with Kroneker symbol on
right-hand side). Returning to (27) and (26) shows that

\begin{equation}
i\int \langle V_{m}^{\ast }\,\widehat{R}V_{k}\rangle dr=i\int \langle
S_{0}\,[V_{k}\,V_{m}^{\ast }]\rangle dr=\delta _{mk}
\end{equation}
This gives the orthogonality rule for the eigenwaves.

The vectors \,$\,V\,\,$\, always can be represented in the form

\begin{equation}
V(r)=e^{i\vartheta (r)}\{a(r)+ib(r)\}\,,\,    a\perp b\,\,,\,  \
a\perp S_{0}\,,   b\perp S_{0}\,,
\end{equation}
where both \,$\,\,a\,\,$\,and\,  \,$\,b\,\,$\, are real-valued vectors situated
in plane \,  \,$\,\Pi (r)\,\,$\, and perpendicular one to another. This
means merely that spin vector draws elliptic trajectory whose main axes are
just \,$\,\,a\,\,$\, and \,$\,\,b\,$\,.\,  \,

2.10. PLANE WAVES AND POLARIZATION DECOMPOSITION.

Consider free MW in an infinite-size uniformly magnetized sample (\,$\,%
S_{0}=const\,$\,). In this envisioned situation, eigenmodes (29) are plane
waves  that is \,$\,\vartheta (r)=\left\langle k,r\right\rangle \,$\,, with \,$\,k\,$\,
being wave vector, and \,$\,\,a\,\,$\, and \,$\,\,b\,$\, are constants. One wave only
corresponds to every \,$\,k\,$\,, because magnetization precession is definitely
clockwise. Putting on \,$\,V_{k}(r)=\,$\, \,$\,V_{k}\exp (i\left\langle k,r\right\rangle
)\,$\,, \,$\,V_{k}=\,$\, \,$\,a_{k}+ib_{k}\,\,$\, in the Eq.26, one easy obtains

\begin{equation}
-\omega _{k}[S_{0}\,b_{k}]+i\omega _{k}[S_{0}\,a_{k}]=\widetilde{W}a_{k}+i%
\widetilde{W}b_{k}\,,\, \, \widetilde{W}\equiv e^{-i\left\langle
k,r\right\rangle }\widehat{\Pi }\widehat{W}\widehat{\Pi }e^{i\left\langle
k,r\right\rangle }
\end{equation}
The vectors \,$\,a\,\,$\,and\,  \,$\,b\,\,$\, always can be such ordered that \,$\,\
\left\langle S_{0}\,[a,b]\right\rangle \,$\, \,$\,>0\,$\,, then (30) yields

\begin{equation}
\omega _{k}p_{k}a_{k}+i\omega _{k}\frac{b_{k}}{p_{k}}=\widetilde{W}a_{k}+i%
\widetilde{W}b_{k}\,,\,  \,p_{k}\equiv \frac{|b_{k}|}{|a_{k}|}
\end{equation}

Since operator \,$\,\widehat{W}\,\,$\,, by its definition, is real-valued
symmetric operator, this equation clearly shows that

\begin{equation}
\widetilde{W}a_{k}=\alpha _{k}a_{k}\,,\,  \widetilde{W}b_{k}=\beta
_{k}b_{k}\,
\end{equation}
Hence, the main axes of MW polarization ellipse, \,$\,a_{k}\,\,$\,and\,  \,$\,%
b_{k}\,$\,, are nothing but eigenvectors of operator (matrix) \,$\,%
\widetilde{W}\,$\,. Besides  two corresponding eigenvalues determine frequency
and eccentricity, \,$\,\,p_{k}\,$\,, of MW:

\begin{equation}
p_{k}=\sqrt{\alpha _{k}/\beta _{k}}\,,\,\,  \,\omega _{k}^{2}=\alpha
_{k}\beta _{k}=\det \,\widetilde{W}\,\,
\end{equation}
(of course, symbol \,$\,\det \,$\,  \,  \,  relates to
2-dimensional projection subspace, while the third eigenvalue is zero).

If the rotation operator \,$\,\widehat{R}\,\,$\, was discarded from Eq.22, instead
of \,$\,-\gamma \widehat{\Pi }\,$\,, this equation would describe monotonic decay
instead of oscillations  with two relaxation rates \,$\,\gamma \alpha _{k}\,$\,, \,$\,%
\gamma \beta _{k}\,$\,. In this sense, \,$\,V_{k}=\,$\, \,$\,a_{k}+ib_{k}\,\,$\, is
decomposition of MW polarization to two relaxation modes.

2.11. SPIN WAVES AND MAGNETOSTATIC WAVES.

According to (22) and (30),

\begin{equation}
\widetilde{W}=\widehat{\Pi }\{W_{0}+\widehat{A}+r_{0}^{2}k^{2}+\widetilde{G}%
(k)\}\widehat{\Pi }\,,
\end{equation}

\begin{equation}
\widetilde{G}(k)\equiv e^{-i\left\langle k,r\right\rangle }\widehat{G}%
e^{i\left\langle k,r\right\rangle }=4\pi \frac{k\otimes k}{k^{2}}\,,
\end{equation}
where \,$\,\widetilde{G}(k)\,$\, is Fourier transform of dipole interaction, and \,$\,%
k^{2}\equiv |k|^{2}\,$\,. To express the frequency more or less evidently, let
us introduce the wave vector projection onto the plain \,$\,\Pi \,$\,,  \,$\,k_{\bot
}\equiv \,$\, \,$\,\widehat{\Pi }k\,\,$\,, the projected anisotropy matrix  \,$\,\widehat{A}%
_{\bot }\equiv \,$\, \,$\,\widehat{\Pi }\widehat{A}\widehat{\Pi }\,$\, and two its
eigenvalues  \,$\,A_{1,2}\,$\,. Then

\begin{equation}
\,\omega _{k}^{2}=(W_{0}+r_{0}^{2}k^{2}+A_{1})(W_{0}+r_{0}^{2}k^{2}+A_{2})+%
\frac{4\pi }{k^{2}}\left\{ k_{\bot
}^{2}(W_{0}+r_{0}^{2}k^{2}+A_{1}+A_{2})-\left\langle k,\widehat{\Pi }%
\widehat{A}\widehat{\Pi }k\right\rangle \right\}
\end{equation}

\[
\alpha _{k}+\beta _{k}=\text{Tr\,  }\widetilde{W}\,\,,\,  \text{%
Tr\,  }\widetilde{W}=W_{0}+r_{0}^{2}k^{2}+A_{1}+A_{2}+4\pi k_{\bot
}^{2}/k^{2}
\]
The latter equality as combined with (32) determines the eccentricity.
Notice that \,$\, k_{\bot }^{2}=\,$\, \,$\,[S_{0}\,k]^{2}\,$\,.

Alternatively, the frequency can be found from the Eqs.25, if exclude
friction and external field and represent the magnetization field in
potential form, \,$\,h_{S}\propto \,$\, \,$\,k\exp (i\left\langle k,r\right\rangle )\,$\,.
This leads to the dispersion equation,

\begin{equation}
\left\langle k,\widehat{\mu }(\omega )k\right\rangle =k^{2}+4\pi
\left\langle k,\{i\omega +\widehat{R}(W_{0}+\widehat{A}+r_{0}^{2}k^{2})%
\}^{-1}\widehat{R}k\right\rangle =0\,,
\end{equation}
whose solution \,$\,\omega =\omega _{k}\,$\, coincides with (36), while wave
polarization follows from Eqs.24.

This consideration fully neglects actual size and shape of a ferromagnet
sample. Dipole interaction in the envisioned boundless MW finds no spatial
scales to compare with its wavelength. That is why dipole contribution to
the dispersion law (36) is insensible to wave length being function of wave
direction only. Perhaps  the waves whose length is ten or more times less
than least from the sample dimensions may be treated in such a way. This MW
are called spin waves (SW). Neither dispersion nor polarization of SW depend
on sample geometry.

However, in relatively long wave dipole interaction inevitably compares
their length with sample dimensions. As the result, dispersion law acquires
the ratio of these two scales:\, \

\begin{equation}
\omega _{k}=\omega (D|k|,k/|k|,r_{0}^{2}k^{2})\,,
\end{equation}
where \,$\,D\,$\, is some characteristic sample size. Waves which for this
correction is essential are called magnetostatic waves (MSW). In most of
modern applications just MSW are directly excited and detected while SW
generated from them in nonlinear wave processes.

{\it REFERENCES}

1. L.D.Landau and E.M.Lifshitz. Electrodynamics of continuous media. Moscow,
Nauka Publ., 1982.

2. A.I.Akhiezer, V.G.Baryakhtar and S.V.Peletminski. Spin waves. Moscow,
Nauka Publ., 1967.

3. E.M.Lifshitz and L.P.Pitaevski. Statistical physics. Part II. Moscow,
Nauka Publ., 1978.

4. Ya.A.Monosov. Nonlinear ferromagnetic resonance. Moscow, Nauka Publ.,
1971.

5. Nonlinear phenomena and chaos in magnetic materials. Editor Ph.E.Wigen.
World Sci. Publ., 1994.

6. V.E.Zakharov, V.S.Lvov and S.S.Starobinetz. Sov.Phys.-Usp., 17, 896
(1975).

7. P.H.Bryant, C.D.Jeffries and K.Nakamura. Spin-wave dynamics in a
ferrimagnetic sphere: experiments and models. In Ref. 5, p. 83.

8. A.N.Slavin, B.A.Kalinikos and N.G.Kovshikov. Spin-wave envelope solitons
in magnetic films. In Ref. 5, p. 209.



\section{Linear waves in films and plates: %
propagator, excitation, and dispersion equation}

\,  \,  In compactly designed microwave devices  current
carrying conductors (wires) are rather suitable inductors and antennas for
magnetic waves (MW) in ferro- and ferri-magnetic films. Therefore, it is
reasonable to unify consideration of the waves and consideration of
electromagnetic impedance of conductors interacting with film. We will
assume characteristic sizes of a circuit be less than \,$\,2\pi c/\omega \,$\,,
with \,$\,\,\omega \,$\, being characteristic operating frequency and \,$\,c\,\,$\, speed
of light. This allows to use quasi-static Maxwell equations.

At the same time, real processes in thin films (thin in the relative sense
that thickness  \,$\,D\,$\,, is much less than length and width) can be
successfully analyzed in the formally infinite film approximation [2,3]. In
particular, domainless static magnetization of a thin film is uniform
everywhere except narrow regions adjoining its edges  with width of order of
film thickness. Concretely, evaluation of dipole force (Eq.2.5, that is Eq.5
in Sec.2) in finite-size plate geometry shows that demagnetizing field
stipulated by film edges decreases at least as \,$\,2D/d\,$\, where \,$\,d\,$\, is distance
from a nearest edge. Thus one can surely suppose \,$\,\,S_{0}=const\,$\,.

Being relatively thin a film may be thick in the absolute sense that \,$\,%
D>>r_{0}\,$\, (\,$\,r_{0}\,$\, is exchange radius). We will consider small-amplitude
(linear) MW excitation in such a film, by wires whose radius much exceeds \,$\,%
r_{0}\,$\,. So thick wires in linear regime can directly excite long
magnetostatic waves (MSW) only. Therefore, it is reasonable to simplify
mathematics by neglecting exchange contribution to the polarizability \,$\,%
\widehat{\chi }\,$\, (see Eq.2.24). This transforms \,$\,\,\widehat{\chi }\,$\, from
integral operator into matrix which locally connects spin precession with
magnetic field.

Under these conditions  general formula (2.38) for MSW frequency takes the
form

\begin{equation}
\omega _{k}=\omega _{N}(D|k|,k/|k|)\,,
\end{equation}
where \,$\,k\,$\, is in-plane wave vector and integer \,$\,N\,$\, enumerates wave branches
different with respect to thickness. Importantly, it shows that (at least at
given wave direction) the group velocity of MSW, \,$\,v_{g}=\,$\, \,$\,\partial \omega
_{k}/\partial k\,$\, \,$\,\propto D\,$\,, is as small as film thickness.\,
\,

3.1. WAVE EXCITATION BY EXTERNAL CURRENTS.

The external field \,$\,h(r,t)\,$\, (see Sec.2) produced by currents is solution to
the equations

\begin{equation}
h(r,t)=\sum h_{m}(r)I_{m}(t)\,\,,\,   \text{{\bf div}\,  }%
\,h_{n}=0\,,\,  \text{{\bf curl}\,  }\,h_{n}=\frac{4\pi }{c}%
J_{n}(r)\,,\,\
\end{equation}
where \,$\,I_{n}\,$\,, \,$\,J_{n}\,$\, and \,$\,h_{n}I_{n}\,$\, are total current, its density
distribution (normalized to unit) and its field for \,$\,n\,$\,-th wire,
respectively. Clearly, so defined \,$\,h_{n}\,$\, are nothing but Green functions
which determine simultaneously electric moving force (EMF) induced in
conductors by time varying film magnetization. The expression for EMF, \,$\,%
\varepsilon _{n}\,$\,, follows merely from Eq.2.14 as the consequence of energy
balance and reads

\begin{equation}
\varepsilon _{n}=\int \langle h_{n}\,\frac{dS}{dt}\rangle\, dr
\end{equation}
This formula is valid for arbitrary strong nonlinear excitation. For weak
excitation to be considered, EMF is linear function of currents  and relation

\begin{equation}
\varepsilon _{n}=\sum \widehat{Z}_{nm}I_{m}
\end{equation}
defines mutual impedances of the wires  \,$\,\widehat{Z}_{nm}\,$\,.

In linear regime we may treat dipole interaction by means of Eqs.2.23-2.25.
Both external and magnetization induced field can be represented in
potential form,

\[
h_{n}=-\nabla U_{n}\,,\,   h_{S}=\sum \widehat{h}_{Sm}I_{m}\,\,  \,%
\widehat{h}_{Sn}=-\nabla \widehat{U}_{Sn}\,\, \
\]
excluding (for external field) places occupied by currents. Here \,$\,\widehat{U}%
_{Sn}\,$\, is contribution obliged to \,$\,n\,$\,-th inductor. The hat means that it is
time convolution operator. Suppose film situated in region\,$\,\,-D/2<\,$\, \,$\,z<D/2\,\,$\,
in parallel to \,$\,xy\,$\,-plane, and expand variables and patterns into Fourier
series:

\[
I_{n}(t)=\int e^{-i\omega t}\widetilde{I}_{n}(\omega )\,  %
\frac{d\omega }{2\pi } %
\,\,,\, U_{n}(r)=\int \exp (ik_{x}x+ik_{y}y) %
\widetilde{U}_{n}(k,z)\, dk\,\,, \,\,\,\
 dk\equiv \frac{dk_{x}dk_{y}}{(2\pi )^{2}}\,\,,
\]
and so on, where \,$\,k=\{k_{x}\,k_{y}\}\,$\, is in-plane wave vector. The potentials
strictly determine magnetization:

\[
\widetilde{S}_{\bot }(\omega,k,z)=-\widehat{\chi }(\omega )\nabla \sum
\widetilde{I}_{m}(\omega )[\widetilde{U}_{Sm}(\omega,k,z)+\widetilde{U}%
\,_{m}(k,z)]\,,
\]
where, after Fourier transform is made, \,$\,\nabla =\,$\, \,$\,\{ik_{x}\,ik_{y}\,\nabla
_{z}\}\,$\,.

3.2. SOURCE FORM-FACTORS.

In the film interior \,$\,\,\nabla ^{2}U_{n}=0\,\,$\,, therefore external
potentials possess simple exponential behavior:

\begin{equation}
\widetilde{U}_{n}(k,z)=\Phi _{n}(k)\exp \{|k|(\sigma _{n}z-D/2)\}\,,\, \
 |k|\equiv \sqrt{k_{x}^{2}+k_{y}^{2}}\,\,
\end{equation}
where \,$\,\sigma _{n}=1\,$\, (or \,$\,-1\,$\,) if \,$\,n\,$\,-th inductor is placed above (or
under) film, and the form-factor, \,$\,\Phi _{n}(k)\,$\,, reflects distribution of \,$\,%
n\,$\,-th current. In particular, let \,$\,n\,$\,-th wire has round cross-section and
oriented strictly along \,$\,y\,$\,-axis at position \,$\,x_{n}\,\,$\,, and distance between
its center line and film top or bottom equals to \,$\,\rho _{n}\,$\,, then

\begin{equation}
\Phi _{n}(k)=\frac{4\pi ^{2}}{ick_{x}}\exp (-|k|\rho _{n}-ik_{x}x_{n})\delta
(k_{y})\,,
\end{equation}
with \,$\,\delta ()\,$\, being Dirac delta-function.

3.3. IMPEDANCE TO MAGNETIC POTENTIAL RELATION.

Besides  the equality \,$\,\nabla ^{2}U_{n}=0\,\,$\, as combined with usual boundary
conditions (continuity of normal magnetic inductance and tangential filed)
allows to express the impedances through the potentials taken at film top or
film bottom:

\begin{equation}
Z_{nm}\equiv e^{i\omega t}\widehat{Z}_{nm}e^{-i\omega t}=\frac{i\omega }{%
2\pi }\int |k|\widetilde{U}_{Sm}(\omega,k, %
\sigma _{n}D/2)\widetilde{U}%
_{n}(-k,\sigma _{n}D/2) \, dk\,\,\
\end{equation}
(we omit details of calculation). Hence, potentials are taken at the surface
most close to receiving ( \,$\,n\,$\,-th) conductor. More exactly, in view of the
definitions (3) and (4), Eq.8 describes the film contribution to full
impedance (which, of course, has also contribution from direct magnetic
interaction between conductors).\,

3.4. WAVE POTENTIAL AND NORMAL WAVE NUMBERS.

According to Eqs.2.25, we have to solve the equation

\begin{equation}
\left\langle \nabla,\widehat{\mu }\,\nabla \right\rangle (\widetilde{U}_{S}+%
\widetilde{U}\,)=0\,\,,\,  \widehat{\mu }\equiv 1+4\pi \widehat{\chi }%
(\omega )\,,
\end{equation}
addressed to either potential of particular current or total one. Of course,
the unit here (as well as in Eqs.2.21) means nothing but the unit matrix. To
carefully consider this equation, we need in introducing two 3-dimensional
unit vectors

\[
\nu \equiv \{k_{x}/|k|,k_{y}/|k|,0\}\,,\,    \overline{z}\equiv
\{0,0,1\}\,,
\]
and \,$\,2\times 2-\,$\,matrix
\[
M=\left[
\begin{array}{cc}
\mu _{\nu \nu } & \mu _{\nu z} \\
\mu _{z\nu } & \mu _{zz}
\end{array}
\right] \equiv \left[
\begin{array}{cc}
\left\langle \nu,\widehat{\mu }\nu \right\rangle & \left\langle \nu,%
\widehat{\mu }\overline{z}\right\rangle \\
\left\langle z,\widehat{\mu }\nu \right\rangle & \left\langle \overline{z}\,%
\widehat{\mu }\overline{z}\right\rangle
\end{array}
\right]
\]
The solution to Eq.9 is

\begin{equation}
\widetilde{U}_{S}+\widetilde{U}=u_{+}\exp (\lambda _{+}|k|z)+u_{-}\exp
(\lambda _{-}|k|z)\,,
\end{equation}
where \,$\,\lambda _{\pm }=\,$\, \,$\,\lambda _{\pm }(\omega )\,$\, are two roots of the
dispersion equation which follows from Eq.9:

\begin{equation}
\mu _{zz}\lambda ^{2}+i(\mu _{\nu z}+\mu _{z\nu })\lambda -\mu _{\nu \nu
}=0\,\,,
\end{equation}
\begin{equation}
\lambda =\lambda _{\pm }\equiv \lambda _{0}\pm \Lambda \,,\, \,\
\lambda _{0}=-\frac{i(\mu _{\nu z}+\mu _{z\nu })}{2\mu _{zz}}\,,\,\,\
\Lambda =\left\{ \frac{\mu _{\nu \nu }}{\mu _{zz}}-\left( \frac{\mu _{\nu
z}+\mu _{z\nu }}{2\mu _{zz}}\right) ^{2}\right\} ^{1/2}\,
\end{equation}
In general \,$\,\lambda _{\pm }\,$\, can be either real or imaginary or complex.

Obviously, matrix \,$\,M\,$\, and hence \,$\,\lambda _{\pm }\,$\, are sensitive to direction
of the in-plane wave vector \,$\,k\,\,$\, but indifferent to its absolute value, \,$\,%
|k| \,$\,. Therefore, as the Eq.9 shows  potentials and magnetization (5) vary
in \,$\,z \,$\,-direction as quickly or slowly as in plane. This is characteristic
property of magnetostatic waves (see Sec.2).

To find the coefficients \,$\,u_{j}\,$\, and then boundary values of potentials  \,$\,%
\widetilde{U}_{Sn}(\omega,k,\pm D/2)\,$\,, for substitution into Eq.8, we must
once again use standard boundary conditions. The final result for the
boundary potential, at the side closest to a given inductor, is presented by
formulas

\begin{equation}
\widetilde{U}_{Sn}(\omega,k,\sigma _{n}D/2)=\Phi _{n}(k)F(\omega,k) \,,
\end{equation}

\begin{equation}
F(\omega,k)\equiv \frac{\lbrack 1-\Delta -i(\mu _{z\nu }-\mu _{\nu
z})]\sinh (\Lambda |k|D)}{(1+\Delta )\sinh (\Lambda |k|D)+2\mu _{zz}\Lambda
\cosh (\Lambda |k|D)}\, \,,
\end{equation}

\begin{equation}
\Delta \equiv \det \,M=\mu _{\nu \nu }\mu _{zz}-\mu _{\nu z}\mu _{z\nu }
\end{equation}

Let us pay attention to the term \,$\,\,i(\mu _{z\nu }-\mu _{\nu z})\,\,$\, in the
nominator of (14). It changes sign when \,$\,k\,\,$\, turns in opposite
direction,\,  \,$\,-k\,\,$\,. This is example of unreciprocity inherent to
phenomena under static magnetization, which leads to difference between  \,$\,%
Z_{nm}\,$\, and \,$\,Z_{mn}\,$\,.\,  \,

3.5. POLARIZABILITY MATRIX.

After excluding exchange from the polarizability \,$\,\widehat{\chi }\,$\, (see
Eq.2.24), the inversion of denominator operator in Eq.2.24 reduces to
algebraic manipulations. The result reads

\begin{equation}
\widehat{\chi }(\omega )=\frac{(\overline{W}_{0}+A_{1}+A_{2})\widehat{\Pi }-%
\widehat{A}_{\bot }+i\widetilde{\omega }\widehat{R}}{(\overline{W}%
_{0}+A_{1})(\overline{W}_{0}+A_{2})-\widetilde{\omega }^{2}}\,,\,\,\
\overline{W}_{0}\equiv W_{0}-i\gamma \widetilde{\omega }\,,   \widetilde{%
\omega }\equiv \frac{\omega }{1+\gamma ^{2}}\,,
\end{equation}
with \,$\,A_{1,2}\,$\, being two nonzero eigenvalues of the projected anisotropy
matrix \,$\,\,\widehat{A}_{\bot }=\,$\, \,$\,\widehat{\Pi }\widehat{A}\widehat{\Pi }\,$\,
(see Sec.2). Additional simplification is implied by the inequality \,$\,\gamma
<<1\,$\, which is always hoped in practice and, indeed, it is very well
satisfied in ferrites ( \,$\,\gamma =10^{-4}\div 10^{-3}\,$\,). By this reason one
can discard all the terms terms containing \,$\,\gamma ^{2}\,$\, (all the more
higher powers), in particular, make no difference between  \,$\,\widetilde{%
\omega }\,$\, and \,$\,\,\omega \,$\,.

3.6. SUSCEPTIBILITY MATRIX.

To detail the susceptibility matrix \,$\,M\,$\,, a few definitions are necessary.
Let \,$\,\,\theta \,$\, be the angle between static magnetization vector \,$\,S_{0}\,$\,
and \,$\,z\,$\,-axis orthogonal to film, and \,$\,\,\varphi \,$\, the angle clockwise
counted between \,$\,y\,$\,-axis and projection of \,$\,S_{0}\,$\, onto the film plane \,$\,x\,$\,-\,$\,%
y \,$\, (so that \,$\,\theta =\,$\, \,$\,\pi /2\,$\, and \,$\,\varphi =0\,$\, correspond to in-plane
magnetization strictly parallel to \,$\,y\,$\,-axis). Next, involve to consideration
two unit-length eigenvectors of \,$\,\widehat{A}_{\bot }\,$\,, \,$\,\overline{A}%
_{1,2}\,\,$\,(of course, mutually orthogonal), which lie in the plane \,$\,\Pi
\bot S_{0}\,$\, and correspond to the eigenvalues \,$\,A_{1,2}\,$\,. We always can
order them so that \,$\,[\overline{A}_{1}\,\overline{A}_{2}]=S_{0}\,$\,. Then
consider the plane, \,$\,z\,$\,-\,$\,S_{0}\,$\,, what passes through both \,$\,z\,$\,-axis and \,$\,%
S_{0}\,$\,, and define one more angle, \,$\,\psi \,$\,, be the angle clockwise counted
between this plane and \,$\,\overline{A}_{1}\,$\,. Thus  \,$\,\psi =0\,$\, means that vector
\,$\,\overline{A}_{2}\,$\, lies in plane \,$\,x\,$\,-\,$\,y\,$\,. Also we need in the quantities

\begin{equation}
\nu _{\Vert }\equiv \nu _{x}\sin \,\varphi +\nu _{y}\cos \,\varphi
=(k_{x}\sin \,\varphi +k_{y}\cos \,\varphi )/|k|=\cos \,\phi \,,
\end{equation}

\[
\nu _{\bot }\equiv \nu _{x}\cos \,\varphi -\nu _{y}\sin \,\varphi
=(k_{x}\cos \,\varphi -k_{y}\sin \,\varphi )/|k|=\sin \,\phi \,\,,
\]
which represent cosine and sine, respectively, of the angle, \,$\,\phi \,$\,,
between the in-plane wave vector \,$\,k\,$\, and projection of \,$\,S_{0}\,$\, onto plane \,$\,x\,$\,%
-\,$\,y\,$\, (i.e. between \,$\,k\,$\, and \,$\,z\,$\,-\,$\,S_{0}\,$\, plane).

In addition, it is convenient to use the notations

\begin{equation}
A_{+}\equiv (A_{1}+A_{2})/2\,\,,\,  A_{-}\equiv (A_{2}-A_{1})/2\,,\; \
W_{1,2}\equiv \overline{W}_{0}+A_{1,2}\,,\,   \omega _{0}^{2}\equiv
W_{1}W_{2}\,,
\end{equation}

\begin{equation}
\widehat{\Omega }=\widehat{\Omega }(\omega )\equiv (\overline{W}%
_{0}+A_{1}+A_{2})\widehat{\Pi }-\widehat{A}_{\bot }+i\omega \widehat{R}
\end{equation}
Recall that at given external field, the vectors \,$\,S_{0}\,$\,, \,$\,\overline{A}%
_{1,2}\,$\, and scalars \,$\,W_{0}\,$\,, \,$\,A_{1,2}\,$\, are completely determined by
solution to the static magnetization equation (2.8). In these notations  \,$\,%
\,A_{1}=\,$\, \,$\,A_{+}-A_{-}\,$\,, \,$\,A_{2}=\,$\, \,$\,A_{+}+A_{-}\,$\,,  and

\begin{equation}
\mu _{zz}=1+\frac{4\pi \Omega _{zz}}{\omega _{0}^{2}-\omega ^{2}}\,,\, \mu
_{\nu \nu }=1+\frac{4\pi \Omega _{\nu \nu }}{\omega _{0}^{2}-\omega ^{2}}%
\,\,,\,\,\mu _{z\nu }=\frac{4\pi \Omega _{z\nu }}{\omega _{0}^{2}-\omega
^{2}}\,,\, \mu _{\nu z}=\frac{4\pi \Omega _{\nu z}}{\omega
_{0}^{2}-\omega ^{2}}\,\,,
\end{equation}
where the four matrix elements of \,$\,\widehat{\Omega }\,$\,  are defined
quite similarly to elements of matrix \,$\,M\,$\,,

\begin{equation}
\Omega _{zz}\equiv \left\langle z,\widehat{\Omega }z\right\rangle
\,,\,\,\Omega _{\nu \nu }\equiv \left\langle \nu,\widehat{\Omega }\nu
\right\rangle \,,\, \Omega _{z\nu }\equiv \left\langle z,\widehat{\Omega }%
\nu \right\rangle \,,\, \Omega _{\nu z}\equiv \left\langle \nu,\widehat{%
\Omega }z\right\rangle
\end{equation}
After a lot of algebra one can find:

\begin{equation}
\Omega _{zz}=(\overline{W}_{0}+A_{+}+A_{-}\cos \,2\psi )\sin ^{2}\theta \,,
\end{equation}

\begin{equation}
\Omega _{\nu \nu }=(\overline{W}_{0}+A_{+})(\nu _{\bot }^{2}+\nu _{\Vert
}^{2}\cos ^{2}\theta )+A_{-}\{(\nu _{\Vert }^{2}\cos ^{2}\theta -\nu _{\bot
}^{2})\cos \,2\psi -2\nu _{\Vert }\nu _{\bot }\sin \,2\psi \cos \,\theta
\}\,,
\end{equation}

\begin{equation}
\Omega _{z\nu }=\Omega _{\times }-i\omega \nu _{\bot }\sin \,\theta \,,\,\
  \Omega _{\nu z}=\Omega _{\times }+i\omega \nu _{\bot }\sin \,\theta \,,
\end{equation}

\begin{equation}
\Omega _{\times }\equiv \{A_{-}\nu _{\bot }\sin \,2\psi -\nu _{\Vert }(%
\overline{W}_{0}+A_{+}+A_{-}\cos \,2\psi )\cos \,\theta \}\sin \,\theta \,
\end{equation}
Importantly, these complicated expressions always compensate one another so
that the determinant (15) has the simple pole only, see below.

3.7. RESPONSE FUNCTION.

To comfortably express the determinant (15), the roots (12) of dispersion
equation (11), and the whole function (14) which connects by Eq.15 the
form-factor of current distribution, \,$\,\Phi _{n}(k)\,$\,, and magnetic potential
of self-induced field, let us introduce the characteristic frequencies  \,$\,%
\,\omega _{u}\,$\,, \,$\,\omega _{1}\,$\,, \,$\,\omega _{2}\,$\, and \,$\,\omega _{3}\,$\,, as
follows:

\begin{equation}
\,\omega _{u}^{2}\equiv W_{1}W_{2}+4\pi (\overline{W}_{0}+A_{+}+A_{-}\cos
\,2\psi )\sin ^{2}\theta \,,
\end{equation}

\begin{equation}
\omega _{1,2}^{2}\equiv W_{1}W_{2}+2\pi (\Omega _{zz}+\Omega _{\nu \nu })\mp
2\pi \sqrt{(\Omega _{zz}+\Omega _{\nu \nu })^{2}-(2\nu _{\bot }\sin \,\theta
)^{2}W_{1}W_{2}}\,,
\end{equation}

\begin{equation}
\omega _{3}^{2}\equiv W_{1}W_{2}+2\pi (\Omega _{zz}+\Omega _{\nu \nu
})+(4\pi \nu _{\bot }\sin \,\theta )^{2}/2=\{\omega _{1}^{2}+\omega
_{2}^{2}+(4\pi \nu _{\bot }\sin \,\theta )^{2}\}/2
\end{equation}
After one more portion of algebra, eventually we obtain:

\begin{equation}
\Delta =\frac{2\omega _{3}^{2}-\omega _{0}^{2}-\omega ^{2}}{\omega
_{0}^{2}-\omega ^{2}}\,,\,   \mu _{zz}=\frac{\omega _{u}^{2}-\omega ^{2}%
}{\omega _{0}^{2}-\omega ^{2}}\,,\,   \mu _{z\nu }-\mu _{\nu z}=-%
\frac{8\pi i\omega \nu _{\bot }\sin \,\theta }{\omega _{0}^{2}-\omega ^{2}}%
\,\,,
\end{equation}

\begin{equation}
\lambda _{0}=\frac{4\pi i\Omega _{\times }}{\omega ^{2}-\omega _{u}^{2}}\,\
,\,  \Lambda =\Lambda (\omega )\equiv \frac{\sqrt{(\omega
_{1}^{2}-\omega ^{2})(\omega _{2}^{2}-\omega ^{2})}}{\omega ^{2}-\omega
_{u}^{2}}
\end{equation}
In this designations  the response function (14) takes the form

\begin{equation}
F(\omega,k)\equiv \frac{(\omega _{0}^{2}-\omega _{3}^{2}-4\pi \omega \nu
_{\bot }\sin \,\theta )\sinh \,[\Lambda (\omega )|k|D]}{(\omega
_{3}^{2}-\omega ^{2})\sinh \,[\Lambda (\omega )|k|D]+(\omega _{u}^{2}-\omega
^{2})\Lambda (\omega )\cosh \,[\Lambda (\omega )|k|D]}\,
\end{equation}

Notice that all the above defined frequencies  \,$\,\,\omega _{0}\,$\,, \,$\,\,\omega
_{u}\,$\,, \,$\,\omega _{1}\,$\,, \,$\,\omega _{2}\,$\, and \,$\,\omega _{3}\,$\,, include complex
factor \,$\,\overline{W}_{0}\equiv \,$\, \,$\,W_{0}-i\gamma \omega \,$\,, hence, they
themselves are complex functions of \,$\,\omega \,$\, although with small imaginary
parts (and thus weakly depending on \,$\,\omega \,$\, ). At \,$\,\gamma \rightarrow 0\,$\,
all they become real values which characterize spectrum of free MSW
(eigenwaves). In particular (see Sec.4), \,$\,\,\omega _{u}\,$\, is the frequency of
spatially uniform spin precession in film.

3.8. FILM INDUCED IMPEDANCE OF CONDUCTORS.

As the consequence of Eqs.8 and 13, mutual impedance of two conductors
situated on one and the same hand from film, is given by

\begin{equation}
Z_{nm}=\frac{i\omega }{2\pi }\int |k|\Phi _{n}(-k)\Phi _{m}(k)F(\omega
,k)\, dk\,,
\end{equation}
where, as well as in (8), integration is performed over all the
two-dimensional in-plane wave vectors  while \,$\,F(\omega,k)\,$\, is determined by
Eq.14 and previous listing of matrix \,$\,M\,$\,.

In all the above formulas  the letter \,$\,\,\omega \,$\,  (as well as all the
frequency related designations \,$\,\omega _{u}\,$\,, \,$\,\omega _{k}\,$\,, \,$\,W\,$\,, \,$\,W_{0}\,$\,
, etc.) means dimensionless angular frequency which equals to actual
dimensional angular frequency \,$\, 2\pi f\,\,$\, (or \,$\,2\pi f\,_{u}\,$\,, etc.)
expressed in units of \,$\,1/\tau _{0}=\,$\, \,$\,2\pi gM_{s}\,$\, (see Sec.2). Hence,
correspondence between quantities like \,$\,\,\omega \,$\, and\,  \,$\,f\,\,$\,\
is established by the rule

\begin{equation}
\omega =f\,[\text{GHz}]/f_{0}[\text{GHz}]\,,\,   f_{0}[\text{GHz}%
]\equiv g[\text{GHz}/\text{kOe}]M_{s}[\text{kOe}]\,,
\end{equation}
where \,$\,f\,\,$\, is dimensional frequency, \,$\,g\approx 2.8\,\,$\,GHz\,$\,/\,$\,kOe is
gyromagnetic ratio, \,$\,M_{s}\,$\, is saturation magnetization, and square brackets
enclose physical unit names. For YIG, \,$\,f_{0}\approx 0.39\,[\,$\,GHz\,$\,]\,$\,.

3.9. IMPEDANCE OF STRAIGHT WIRES AND LOOPS.

Substituting (7) in (32), we obtain mutual impedance for a pair of
cylindrical wires which are directed along \,$\,y\,$\,-axis in parallel one to
another and to film and lying on the same hand from it (again we omit rather
tremendous manipulations). To express the result in pleasant form, let us
measure distances and sizes in centimeters and impedances in Ohm. Then

\begin{equation}
\frac{Z_{nm}[\text{Ohm}]}{w[\text{cm}]f\,[\text{GHz}]}=4\pi
i\int_{0}^{\infty }e^{-q(\rho _{n}+\rho _{m})}\frac{[(1-\Delta )\cos
(qx)+(\mu _{z\nu }-\mu _{\nu z})\sin (qx)]\sinh (\Lambda qD)}{(1+\Delta
)\sinh (\Lambda qD)+2\mu _{zz}\Lambda \cosh (\Lambda qD)}\frac{dq}{q}\,
\end{equation}
Here \,$\,x=\,$\, \,$\,x_{n}-x_{m}\,$\, is distance between the wires  and \,$\,w\,\,$\, is their
length (in centimeters). The integral is taken over \,$\, q\equiv \,$\, \,$\,k_{x}>0\,$\,,
\,$\,k_{y}=0\,$\,, that is in all \,$\,\mu \,$\,-related parameters of the integrand it
should be put on\,  \,$\,\,\nu =\,$\, \,$\,\{1,0,0\}\,$\,, \,$\,\nu _{\bot }=\,$\, \,$\,\cos
\,\varphi \,$\,, \,$\,\,\nu _{\Vert }=\,$\, \,$\,\sin \,\varphi \,$\,, while the frequency \,$\,%
\omega \,$\, should be mentioned in accordance with (33).

For the case when two parallel wires are situated on the opposite parties
from the film, evaluation of corresponding boundary potentials yields (in
the same units):
\begin{equation}
\frac{Z_{nm}}{wf\,}=2\pi i\int_{-\infty }^{\infty }e^{-|q|(\rho _{n}+\rho
_{m})+iqx}\left\{ e^{-|q|D}-\frac{2\mu _{zz}\Lambda \exp (\lambda _{0}|q|D)}{%
(1+\Delta )\sinh (\Lambda |q|D)+2\mu _{zz}\Lambda \cosh (\Lambda |q|D)}%
\right\} \frac{dq}{|q|}
\end{equation}
(to be accompanied by definitions (12) and (15)). Of course, here \,$\,\,\nu =\,$\, \,$\,%
\{\,$\,sign\,$\,(q),0,0\}\,$\, in the integrand. Due to the definition (8),  \,$\,Z=\,$\, \,$\,%
R-2\pi ifL\,$\,, where \,$\,R=\,$\,Re \,$\,Z\,\,$\, and \,$\,L=-\,$\,Im \,$\,Z/2\pi f\,\,$\, play roles of
resistance and inductance.

In fact, what we made is evaluation of impedances per unit length for long
line inductors with not taking into account disturbance of their parallelity
and edge effects. Formulas (34) and (35) can be obviously generalized to
strip-shaped wires and to the case when transmitting or/and receiving
inductor consist of two parallel wires which form loop (or many wires which
form antennae lattice) and can perform more or less wave selection. What is
for more complicated configurations  to investigate them one should return
to formulas (13)-(15) and (32).

By its definition, the distance between any wire and film, \,$\,\rho _{n}\,$\,, can
not be less than the wire radius. Closely looking on integrands in (34) and
(35) we see, firstly, that only such the waves are excited (and contributing
to the impedance) whose wavelength notably exceeds \,$\,\rho _{n}+\rho _{m}\,$\,.
Secondly, impedance depends on the ratio \,$\,\delta \equiv \,$\, \,$\,2D/(\rho
_{n}+\rho _{m})\,$\, but not on \,$\,D\,$\, or \,$\,\rho _{n}\,$\, separately.

3.10. RESISTANCE OF PARALLEL WIRE.

Exact analytical integration in (34) is impossible. But in the important
special case when conductors are lasting along external field, \,$\,H_{0}\,$\,,
which lies in the film plane, there exists satisfactory analytical
approximation. In particular, for magnetic contribution to resistance of a
single wire (at \,$\,x=0\,$\, ), under neglecting anisotropy, we found the estimate

\begin{equation}
\frac{R_{11}^{wire}[\text{Ohm}]}{w[\text{cm}]}\approx \frac{2\pi \left( 2\pi
+H_{0}\right) fX(f)}{[1-X^{2}(f)]\,\text{arctanh\,  }X(f)}\exp
\left\{ -\frac{\rho }{D}\,\text{arctanh\,  }X(f)\right\} \,,\,\,%
\text{if\,  }  f_{u}<f<f_{0}(H_{0}+2\pi )\,,
\end{equation}

\[
X(f)\equiv \frac{f^{2}-f_{u}^{2}}{(2\pi f_{0})^{2}}\,,\,  \
f_{u}\equiv f_{0}\sqrt{H_{0}\left( H_{0}+4\pi \right) }\,,\, \, \text{%
arctanh\,  }X=\frac{1}{2}\ln \frac{1+X}{1-X}
\]
In these formulas \,$\,f\,$\, is frequency in GHz, \,$\,H_{0}\,$\, expressed in units of \,$\,%
M_{s}\,$\,, and \,  \,  \,$\,f_{u}\,$\,  is the uniform spin precession
frequency of in-plane magnetized film (with no anisotropy, see next
Section). Outside of the marked frequency interval, resistance turns into
zero. Corresponding wire inductance can be simply estimated at  \,$\,\delta
\lesssim 0.5\,$\, only, and in this case

\begin{equation}
\frac{L_{11}^{wire}[\text{nH}]}{w[\text{cm}]}\approx -\left( 2+H_{0}/\pi
\right) e^{-X(f)}%
\mathop{\rm Ei}%
(X(f)) \,,
\end{equation}
with \,$\,%
\mathop{\rm Ei}%
(\xi )\,\,$\, being exponential integral function [4].

Notice that at in-plane magnetization \,$\,W_{0}=H_{0}\,$\, because in-plane
demagnetization factors are equal to zero.

\begin{figure}
\includegraphics{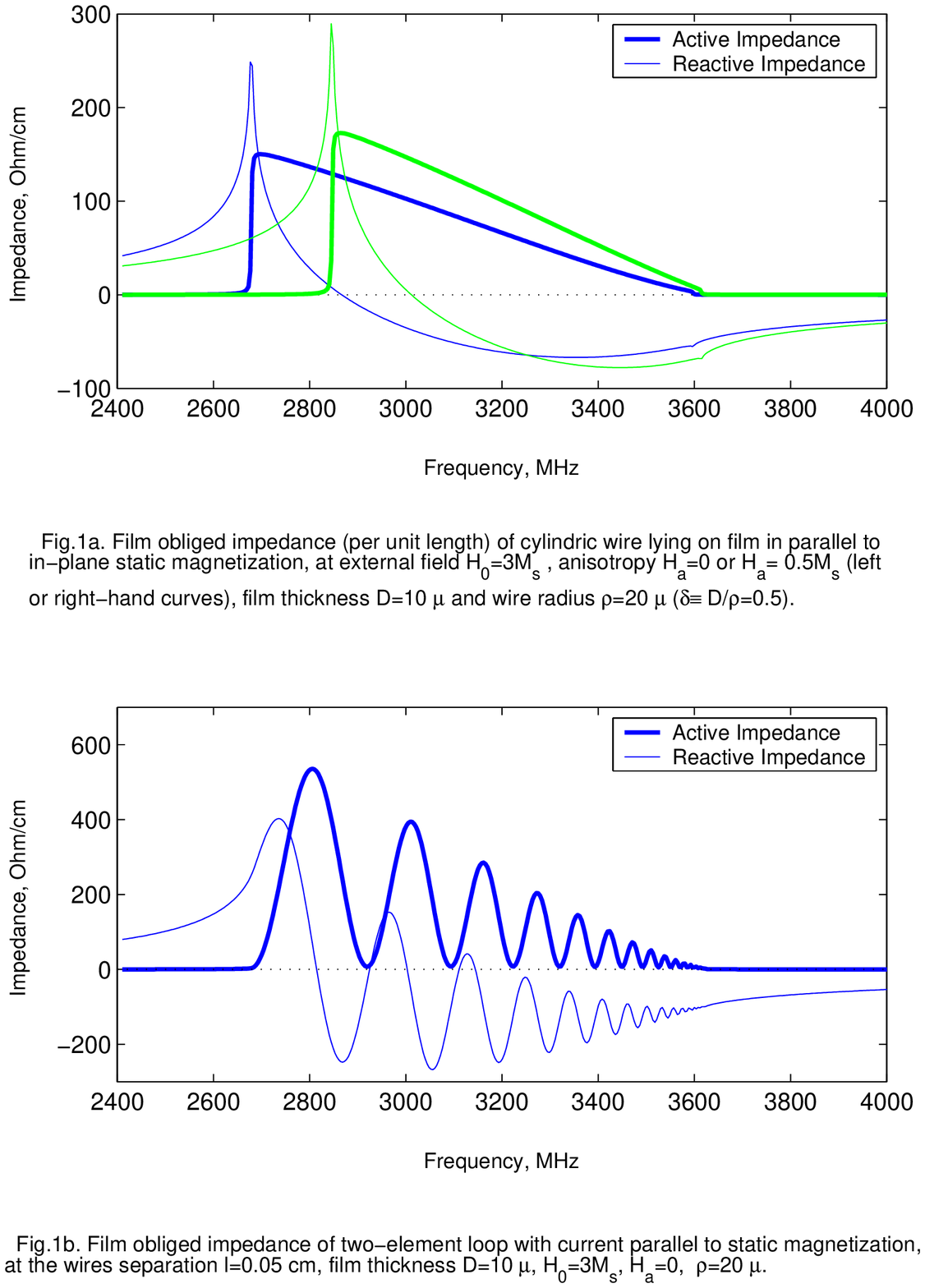}
\end{figure}

\begin{figure}
\includegraphics{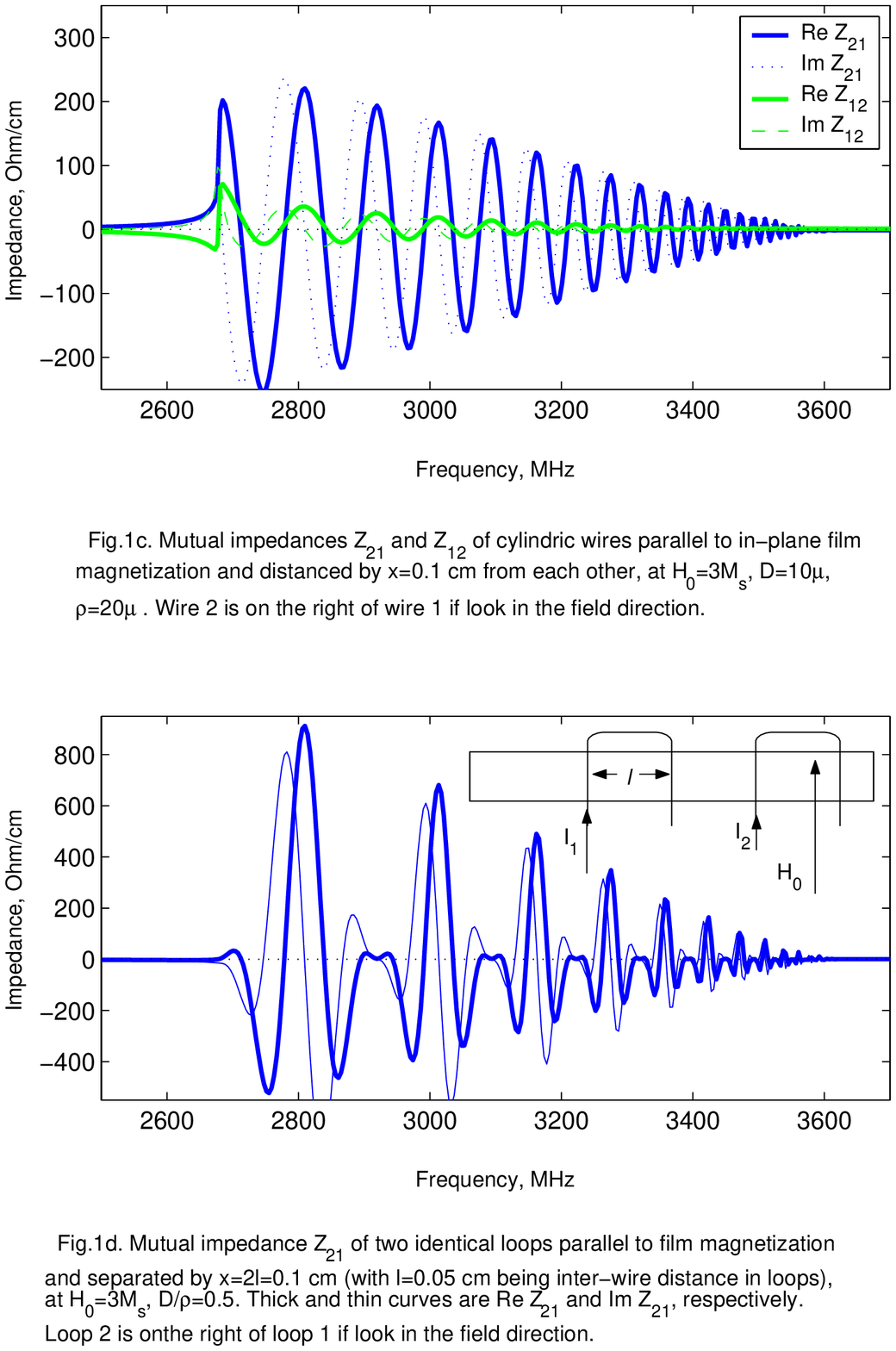}
\end{figure}

3.11. NUMERICAL EXAMPLES AND DAMON-ESHBACH WAVES.

In general, the integrals can be obtained by numerically. For this purpose,
formula (14) is better than (31) because numerical procedure deals directly
with matrix (16). Some results are illustrated by Figs.1a-1d, 2a-2b and
3a-3b. Since we are most interested in YIG samples  the value \,$\,M_{s}\approx
140\,\,$\,Oe was substituted.

The Fig.1a relates to the same case as the Eq.36. To show characteristic
influence by anisotropy, two pairs of curves are presented, for \,$\,A_{1,2}\,$\, \,$\,%
=0 \,$\, and \,$\,\,A_{2}=\,$\, \,$\,-A_{1}=\,$\, \,$\,H_{a}=0.5\,$\,, supposing its main axes are \,$\,x\,$\,-
and \,$\,z\,$\,-axes at \,$\,y\,$\,-directed field (i.e. \,$\,\psi =0\,$\,). As typically,
anisotropy increases frequencies of MSW running across the field.
Unfortunately, real\,  anisotropy in YIG (function \,$\,A(S)\,$\, in Eq.2.2)
involves essential complications. For the present, in next examples it was
neglected at all. Inductors lie on one and the same top side of film and
last along \,$\,y\,$\,-axis  while \,$\,H_{0}\,$\, is oriented either along \,$\,y\,$\,-\,
or \,$\,x\,$\,-axis.

The Fig.1b presents impedance for the loop consisting of two parallel
wires which continue one another and carry the same current but in
opposite directions (see inset in Fig.1d). Any inductor parallel to
magnetization naturally generates so-called Damon-Eshbach
magnetostatic waves first discovered in [5]. They run perpendicular
to \,$\,S_{0}\,$\,. Uniquely, their dispersion law can be found (if neglect
anisotropy) in simple analytical form (see Sec.4):
\begin{equation}
\omega _{k}=\omega _{DE}(D|k|)\equiv \sqrt{H_{0}\left( H_{0}+4\pi \right)
+4\pi ^{2}\{1-\exp (-2D|k|)\}}\,
\end{equation}
Maximums of the loop resistance on Fig.1b exactly correspond to
Damon-Eshbach (DE) waves with lengths \,$\,2ml\,$\,, where \,$\,l\,$\, is the width of the
loop (i.e. inter-wire distance) and \,$\,m=1,3,...\,$\, odd integers  just as one
could expect. Notice that all the spectrum of DE waves lies above the
uniform precession frequency, \,$\,\omega _{u}=\,$\, \,$\,\sqrt{H_{0}\left( H_{0}+4\pi
\right) }\,$\,. In dimensional form, at \,$\,H_{0}/M_{s}=3\,$\, chosen for these
examples  \,$\,f_{u}\approx 2.68\,\,$\,GHz.

Figs.1c and 1d show mutual impedances of two wires and two loops
respectively, under the same orientation. Both ''from left to right''
impedance and ''from right to left'' are presented in Fig.1c. The latter
clearly demonstrates violation of the reciprocity: every inductor placed
above the film top better excites waves going clockwise from \,$\,H_{0}\,$\, and \,$\,%
S_{0}\,$\, than inverse waves. To change preferred direction one must remove
inductors under film. The half of the sum of these two impedances  by its
magnitude, equals approximately to impedance on Fig.1a. One may find also
that maximums of absolute value of impedance in Fig.1d and maximums of
resistance in Fig.1b take place at the same frequencies and have equal
amplitude ratios. In other words  loops separation strongly influences phase
of mutual impedance but slightly its frequency filtering characteristics.

Figs.2a and 2b relate to wires and loops which are oriented perpendicular to
magnetization (see inset in Fig.2b) and hence excite MSW running along it.
The essential difference from previous case is that these MSW are irradiated
symmetrically (reciprocity takes place), and their spectrum lies
below\,  \,$\,\omega _{u}\,$\,. Since this highest frequency responds to
least wave number, the group velocity of these waves is negative, i.e.
directed in opposite to phase velocity.

\begin{figure}
\includegraphics{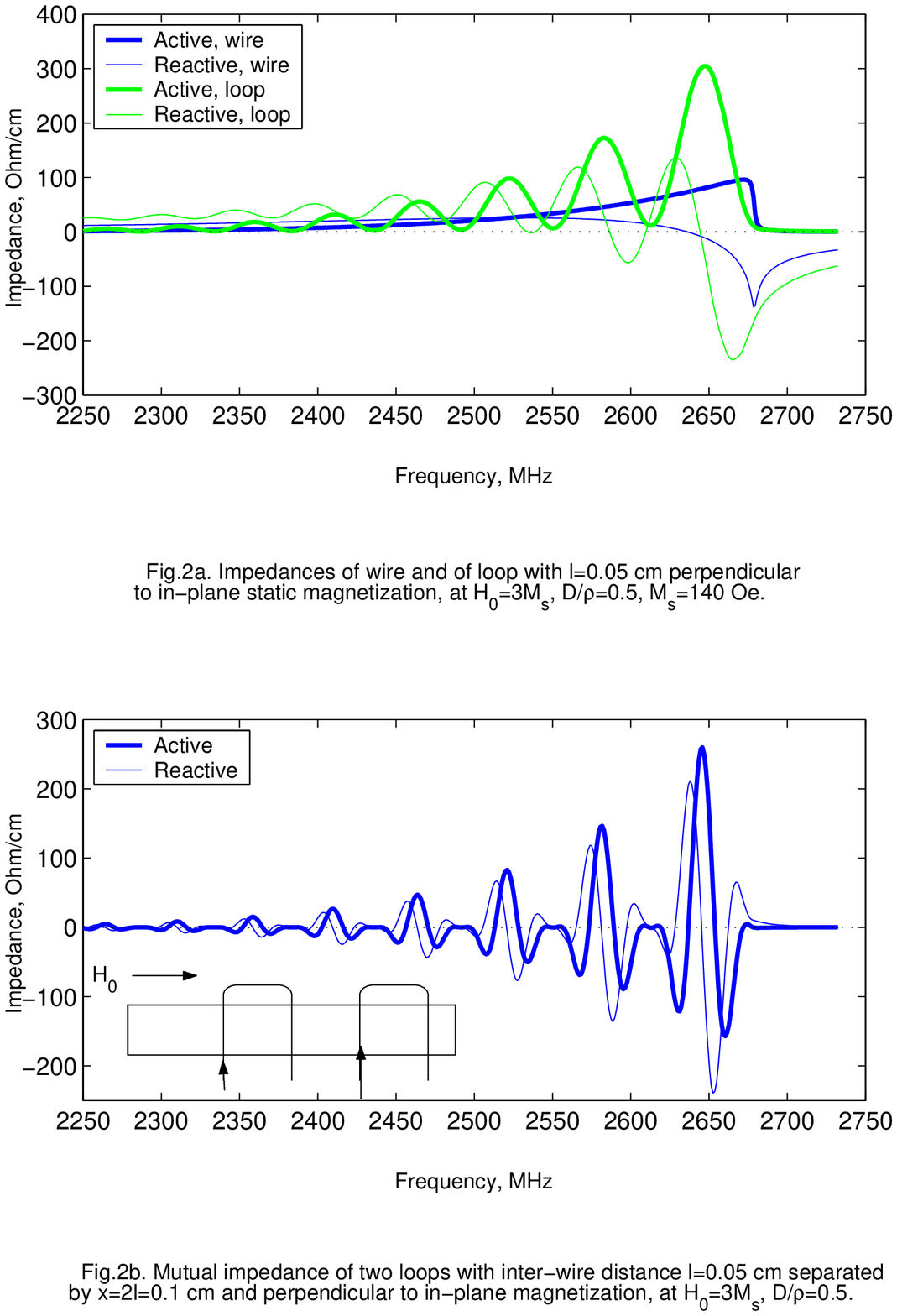}
\end{figure}

\begin{verbatim}
\end{verbatim}

3.12. ROLE OF FILM THICKNESS.

Let us return to wire parallel to magnetization, and consider Fig.3a which
contains the series of resistance via frequency curves for different
thickness values. Naively, one would predict nearly linear dependence \,$\,%
R(D)\, \,$\, if suppose the EMF (voltage) be proportional to time-varying
magnetization, \,$\,S_{\bot }\,$\,, and to a number of contributing spins (that is
to thickness), \,$\,\varepsilon \propto \,$\, \,$\,D|S_{\bot }|\,$\,, while \,$\,S_{\bot }\,$\,
proportional to exciting current, \,$\,|S_{\bot }|\,$\, \,$\,\propto I\,$\,. But
Figs.3a,3b and formula (36) show that, at better excited lower part of the
spectrum, resistance is almost independent on \,$\,D\,\,$\, (higher frequencies are
rejected merely by the exponential wire form-factor, see Eq.7).

What is the matter? The answer comes from Eqs.1 and 38: the smaller
thickness the smaller group velocity of irradiated waves  \,$\,v_{g}\,$\,,
therefore, the smaller is energy outflow from the inductor. Assume this
outflow be proportional to product \,$\,Dv_{g}|S_{\bot }|^{2}\,$\,, equate it to
the pumped power \,$\,RI^{2}\,$\, and combine the resulting relation with \,$\,%
\varepsilon \propto \,$\, \,$\,D|S_{\bot }|\,$\,. These reasoning yield \,$\,R\propto \,$\, \,$\,%
D/v_{g}\,$\,. Taking into account that the ratio \,$\,\,v_{g}/D\,$\,  is a function of
product the \,$\,D|k|\,$\, or, equivalently, of the frequency only, we get the
explanation of approximate constancy of resistance.

From Fig.3a and Eq.36 it is evident that at \,$\,D\gtrsim \,$\, \,$\,1.5\rho \,$\, the
form-factor becomes unimportant and resistance almost independent on \,$\,D\,\,$\,
in all the frequency region. The strong rise of resistance at frequencies
close to\,  \,$\,f_{u}+\,$\, \,$\,2\pi f_{0}\,$\, reflects fast falling of group
velocity in this region. Indeed, the Eq.38 implies for DE waves

\begin{equation}
v_{g}=D\{(H_{0}+2\pi )^{2}-\omega ^{2}\}/\omega \,
\end{equation}
Hence, the assumption \,$\,R\propto \,$\, \,$\,D/v_{g}\,$\, implies \,$\,R\propto \,$\, \,$\,%
1/(H_{0}+2\pi -f/f_{0})\,$\, what is qualitatively confirmed by both Eq.36 and
numerical results. This rise becomes better clear if notice that less group
velocity means greater density of (excited) states (what is highlighted by
right-hand peaks condensation in Fig.1b).

Incidentally, we can conclude that \,$\,|S_{\bot }|\propto I/D\,$\,, i.e. at given
current the more thin is film the stronger swing of magnetization and thus
the closer nonlinear excitation regime. But EMF voltage signal remains
approximately the same, even if  \,$\,|S_{\bot }|\,$\, is comparable with unit.
Therefore, to get greater voltage signal in nonlinear regime, one is
enforced to make film thicker.

\begin{figure}
\includegraphics{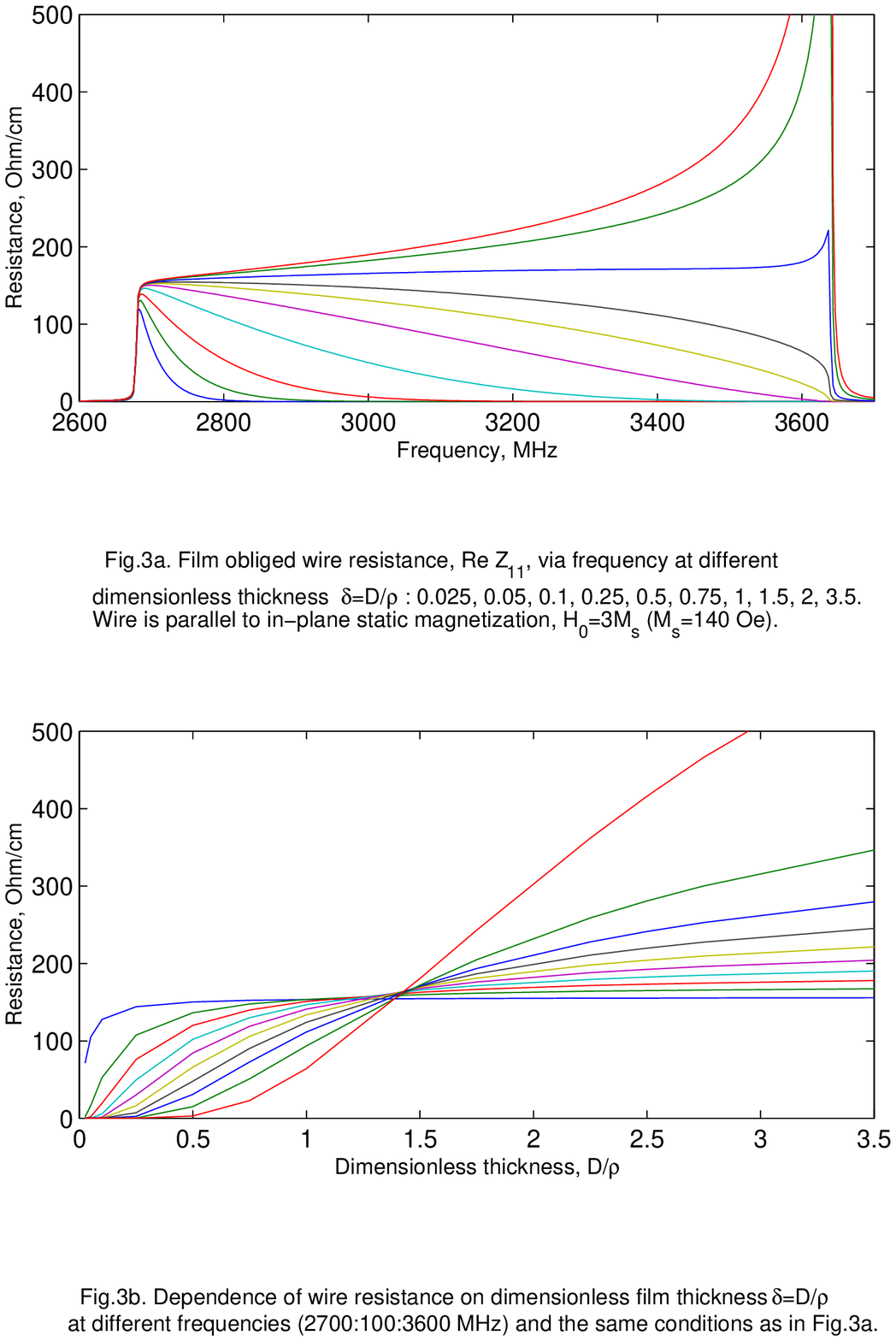}
\end{figure}

3.13. ROLE OF FRICTION.

According to Eqs.16 the factual dimensionless friction coefficient is not \,$\,%
\gamma \,$\, itself but product \,$\,\,\Gamma \equiv \gamma \omega \,$\, (also small
quantity). In dimensional form,

\begin{equation}
\Gamma =\gamma \omega /\tau _{0}=2\pi \gamma f\lesssim 2\pi \gamma
f_{0}(H_{0}+2\pi )\,
\end{equation}
The latter estimate relates to in-plane magnetized film where, as we could
conclude, \,$\,f_{0}(H_{0}+2\pi )\,$\, is the upper bound of MSW spectrum (for small
anisotropy and long MSW as compared with \,$\,r_{0}\,$\, ; see Sec.4).

In infinite-size film under above formal consideration any external source
generates continuous wave spectrum, regardless of concrete \,$\,\Gamma \,$\, value,
i.e. non-resonant excitation takes place. That is why \,$\,\gamma \,$\, in no way
manifests itself in (36). But real finite-area film has discrete MSW
spectrum. If characteristic frequency separation of excitable eigenwaves  \,$\,%
\delta f\,$\,, essentially exceeds their spectral broadening, \,$\,\delta f\,$\, \,$\,>>\,$\, \,$\,%
\Gamma /2\pi \,$\,, then it is principally possible to resonantly distinguish
them.

Let us allow that we select only one-dimensional set of DE modes running in \,$\,%
x\,$\,-direction and uniform in \,$\,y\,$\,-direction (\,$\,\,k_{y}=0\,\,$\,). Their separation
by wavenumber is on order of  \,$\,\approx 2\pi /d\,$\,, where \,$\,d\,$\, stands for wire
length. Hence, frequency separation is \,$\,\delta f\approx \,$\, \,$\,f_{0}(\partial
\omega _{k}/\partial k)\,$\, \,$\,\cdot (2\pi /d)\,$\, \,$\,=\,$\, \,$\,v_{g}/d\,$\, (with \,$\,v_{g}\,$\, being
dimensional group velocity), and the resonance is possible if \,$\, \Gamma
d\precsim v_{g}\,$\,. Expressing \,$\,\,v_{g}\,$\, from Eq.38, for example, at \,$\,%
H_{0}\sim 3\,$\, and \,$\,D\sim 10\,[\mu ]\,\,$\,, we obtain \,$\,\,v_{g}\approx \,$\, \,$\,4\pi
^{2}D/\tau _{0}\omega _{u}\,$\, \,$\,\sim \,$\, \,$\,10^{7}\,\,$\,[cm/s]. For\,  \,$\,\gamma
\sim \,$\, \,$\,3\cdot 10^{-4}\,$\, and \,$\,D\sim \,$\, \,$\,10\,[\mu ]\,\,$\,, any length \,$\,d\,$\, \,$\,<<\,$\, \,$\,%
10\,\,$\,  \lbrack cm] occurs sufficiently small!

In such the case is the theory applicable to real films with\,  \,$\,%
d\sim 0.5\,\,$\,[cm]? Yes  if wave selection, under realistic source
form-factor, is not so perfect as was assumed. Even if a few modes only with
nonzero \,$\,k_{y}\,$\, \,$\,\sim \,$\, \,$\,2\pi /w\,\,$\, are excited in addition to \,$\,k_{y}\,$\, \,$\,=0\,$\,
, the sufficiently small length easy fails down to 0.1 [cm] or less.
Besides  in real finite-amplitude process directly excited modes transmit
their energy to other modes by means of non-linear wave interactions. Thus
the latters effectively increase friction and approach situation to the
idealized model.

{\it REFERENCES}

1. A.I.Akhiezer, V.G.Baryakhtar and S.V.Peletminski. Spin waves. Moscow,
Nauka Publ., 1967.

2. Nonlinear phenomena and chaos in magnetic materials. Editor Ph.E.Wigen.
World Sci. Publ., 1994.

3. A.N.Slavin, B.A.Kalinikos and N.G.Kovshikov. In Ref. 2, p. 209.

4. M.Abramowitz and I. A. Stegun. Handbook of Mathematical Functions.
Chapter 5. N-Y, Dover Publ., 1965.

5. R.W.Damon and J.R.Eshbach. J.Phys.Chem.Sol., 19 (1961) 308.



\section{Linear waves in films and plates: %
eigen-modes and dispersion laws}

4.1. IN-PLANE WAVE REPRESENTATION.

Consider magnetic field created by the magnetization wave

\[
S_{\bot }\equiv Ve^{-i\omega t}\,,\, V\equiv V(z)\exp
\{i(k_{x}x+k_{y}y)\}\,,
\]
where \,$\,k=\,$\, \,$\,\{k_{x}\,k_{y}\}\,$\, is in-plane wave vector. For any function \,$\,G(r)\,$\,%
\,  \,  \,  whose 3-dimensional Fourier transform is
known be \,$\,\widetilde{G}(K)\,$\,, with \,$\,K=\,$\, \,$\,\{k_{x}\,k_{y}\,k_{z}\}\,\,$\,, and any
function \,$\,\,f(z)\,\,$\,, the relations take place as follows

\[
\int G(r-r^{\prime })f(z^{\prime })e^{i(k_{x}x^{\prime }+ %
k_{y}y^{\prime })}\, dr^{\prime }\, =\,  %
e^{i(k_{x}x+k_{y}y)} \int G(k,z-z^{\prime })f(z^{\prime
})\, dz\,,\, G(k,z)\equiv \int e^{ik_{z}z}\widetilde{G}(K) %
\, \frac{dk_{z}}{2\pi }
\]
If applying this theorem to dipole interaction matrix function defined by
(2.5) and (2.6), in accordance with (2.35) we have
\begin{equation}
G(k,z)=2\pi \left[
\begin{array}{cc}
(k\otimes k)/|k| & i\left[
\begin{array}{c}
k_{x} \\
k_{y}
\end{array}
\right] \,\text{sign}(z) \\
i\left[
\begin{array}{cc}
k_{x} & k_{y}
\end{array}
\right] \,\text{sign}(z) & 2\delta (z)-|k|
\end{array}
\right] \,e^{-|k||z|}\,\,,
\end{equation}
where \,$\,\delta (z)\,$\, is Dirac delta-function. Expression (1) is dipole
interaction kernel in \,$\,(k,z)\,$\,-representation. In this representation,
operator \,$\,\widehat{G}\,$\,, the linear dynamical operator \,$\,\widehat{W}\,$\, defined
in Eq.2.22, and the self-induced field take the form

\begin{equation}
\widehat{G}f\equiv \int_{-D/2}^{D/2}G(k,z-z^{\prime })f(z^{\prime
})dz^{\prime }\,\,  \widehat{W}=W_{0}+\widehat{A}+r_{0}^{2}(|k|^{2}-\nabla
_{z}^{2})+\widehat{G}\,,\,  h_{S}=-\widehat{G}V
\end{equation}

4.2. SINGULARITY OF DIPOLE INTERACTION.

From Eq.1 we see that in the long wave limit, when \,$\,|k|D\,$\, \,$\,\rightarrow 0\,$\,,
all the matrix elements of \,$\,G(k,z)\,$\, turn into zero except the only singular
element, \,$\,G_{zz}\rightarrow \,$\, \,$\,4\pi \delta (z)\,$\,, which performs purely
local connection in \,$\,z\,$\,-direction:

\begin{equation}
\widehat{W}\,\Rightarrow W_{0}+\widehat{A}-r_{0}^{2}\nabla _{z}^{2}+4\pi \,%
\overline{z}\otimes \overline{z}\,,\,   \overline{z}\equiv \{0,0,1\}
\end{equation}
In this limit, \,$\,\,h_{S}(z)\Rightarrow \,$\, \,$\,-4\pi \overline{z}V_{z}(z)\,\,$\,,
that is the field \,  vanishes everywhere outside film while in its
interior in any layer \,$\,z=const\,\,$\, it fully reduces to magnetization of that
layer.

This trivial fact of magnetostatics means  with respect to spin dynamics
that if being uniformly magnetized any separate flat layer acts on itself
only. Therefore, spin oscillations in different layers can behave
independently, resulting in strange property of MSW: at \,$\,|k|D\,$\, \,$\,<<1\,$\, many
MSW branches with different normal wave numbers have almost the same
frequencies very close to the frequency of uniform spin precession, \,$\,\omega
_{u}\,$\,.

Hence, \,$\,\omega _{u}\,$\, is essentially peculiar degenerated point of MSW\
spectrum. Although exchange interaction forbids too arbitrary \,$\,z\,$\,%
-distributions and removes exact degeneracy, the latter remains important in
absolutely thick plates  hindering resonant excitation of too long MSW.

4.3. UNIFORM PRECESSION.

Under uniform precession, all spins in the sample rotate with exactly the
same phase, that is \,$\,V=const\,$\,, and \,$\,\vartheta (r)=const\,$\, in Eq.2.29.
According to (3), in this situation, as in case of boundless MW considered
in Sec.2.10, operator \,$\,\widehat{W}\,\,$\, transforms to algebraic one. As the
consequence, clearly, again the polarization decomposition is possible.
Therefore, we can write

\begin{equation}
\omega _{u}^{2}=\det \,\widetilde{W}\,,\,   \widetilde{W}\equiv \,%
\widehat{\Pi }(W_{0}+\widehat{A}+4\pi \,\overline{z}\otimes \overline{z})%
\widehat{\Pi }
\end{equation}
Since matrix \,$\,\widetilde{W}\,\,$\, has exactly the same structure as matrix in
Eq.2.34, with \,$\,\overline{z}\,\,$\, in place of \,$\,k/|k|\,$\, and zero in place of \,$\,%
r_{0}^{2}k^{2}\,$\,, in fact the answer is already presented by Eq.2.36, namely,

\begin{equation}
\omega _{u}^{2}\equiv (W_{0}+A_{1})(W_{0}+A_{2})+4\pi (W_{0}+A_{1}\sin
^{2}\psi +A_{2}\cos ^{2}\psi )\sin ^{2}\theta
\end{equation}
The angles \,$\,\theta \,$\, and \,$\,\psi \,$\, were introduced in Sec.3.6. This
expression is equivalent to Eq.3.26 obtained when considering MW excitation.

The polarization and eccentricity also can be obtained with recipes of
Sec.2.10 and 2.11. Of course, in absence of anisotropy the main axes look
as\,  \,$\,a\parallel \overline{z}\,$\, and \,$\,\,b\parallel \lbrack S_{0}\,%
\overline{z}]\,$\,. In this case,

\begin{equation}
p=|b|/|a|=\sqrt{1+\frac{4\pi }{W_{0}}\sin ^{2}\theta }\,,\,  \,(A=0\,)\,,
\end{equation}
that is polarization ellipse of uniform precession is always in-plane
stretched (spins do not like piercing the plane).

For further, let us agree that \,$\,A=0\,$\, and \,$\,\psi =0\,$\, will mark neglected (or
effectively negligible) anisotropy or specially oriented anisotropy (see
Sec.3.6), respectively.

4.4. DISPERSION EQUATION.

In fact, the results of Sec.3 are sufficient to analyze all the variety of
free eigenwaves in (absolutely thick) film geometry. As usually, their
frequencies are nothing but poles of the response function, \,$\,F(\omega,k)\,$\,
(see Eqs.3.14, 3.31), at vanishing friction. Hence, to get the dispersion
equation, we must equate denominator of (3.31) to zero. Elimination of
friction from formulas of Sec.3 is achieved merely by returning real
quantity \,$\,W_{0}\,$\, (static effective field) in place of complex one, \,$\,%
\overline{W}_{0}\, \,$\,introduced\,$\, \,$\,in Eq.3.16. Thus we come to the
dispersion equation as follows

\begin{equation}
\coth \,[\Lambda (\omega )|k|D]\, =\frac{(\omega _{3}^{2}-\omega ^{2})\text{%
sign}\,(\omega ^{2}-\omega _{u}^{2})}{\sqrt{(\omega _{1}^{2}-\omega
^{2})(\omega _{2}^{2}-\omega ^{2})}}\,\,,\, \,\Lambda (\omega )=\frac{%
\sqrt{(\omega _{1}^{2}-\omega ^{2})(\omega _{2}^{2}-\omega ^{2})}}{|\omega
^{2}-\omega _{u}^{2}|}\,
\end{equation}
In view of Eqs.3.22-3.28 and Eq.4, here \,$\,\,\omega _{3}=\,$\, \,$\,\omega
_{3}(k/|k|)\,\,$\, is depending on orientation of in-plane wave vector while \,$\,%
\omega _{u}^{2}\,\,$\, is constant.

At given in-plane wave vector, Eq.7 has either one or infinitely many real
roots  with respect to \,$\,\omega ^{2}\,$\,, determining different MSW branches.
For any of roots  \,$\,\omega ^{2}=\,$\, \,$\,\omega _{N}^{2}(k)\,\,$\,, we can obtain also
corresponding pair of out-plane wave numbers  defined in accordance with
Eqs.3.10-3.12 and 3.30 :

\begin{equation}
iq_{\pm }(k)=\lambda _{\pm }(\omega )|k|=ik_{0}(\omega )\pm \Lambda (\omega
)|k|\,,\,  k_{0}(\omega )\equiv \lambda _{0}(\omega )|k|/i=\frac{%
4\pi \Omega _{\times }|k|}{\omega ^{2}-\omega _{u}^{2}}\,,
\end{equation}
with \,$\,\,\Omega _{\times }=\,$\, \,$\,\Omega _{\times }(k/|k|)\,$\, being expressed by
Eq.3.15, and \,$\,\omega =\,$\, \,$\, \omega _{N}\,(k)\,$\,. Notice that \,$\,\,\Omega
_{\times }\,$\, and consequently \,$\,k_{0}(\omega )\,$\, always are real-valued.

4.5. TWO TYPES OF WAVES.

In a bulk wave, by its definition, magnetization harmonically oscillates
along \,$\,z\,$\,-axis  i.e. \,$\,\Lambda (\omega )\,$\, takes some imaginary value.
Clearly, this is the case if a root of Eqs.7 belongs to the interval

\begin{equation}
\omega _{1}^{2}<\omega ^{2}<\omega _{2}^{2}\, \,\
\end{equation}

Otherwise, \,$\,\Lambda \,$\, is real and magnetization varies exponentially
responding to what is usually termed surface wave. However, due to the
common scaleless nature of MW governed by dipole interaction, characteristic
exponents  \,$\,\pm \Lambda (\omega )|k|\,$\,, are of order of in-plane wave
number. At \,$\,D|k|\,$\, \,$\,>>1\,$\,, such a wave is indeed concentrated in the vicinity
of film surfaces. But in practically important case, when in-plane
wavelength is greater than thickness  this wave is indistinguishable from
bulk one. The Figs.1a-1d and 3a-3b relate just to such ''surface-balk'' MSW
(see below).

At arbitrary wave vector \,$\,k\,$\,, Eq.7 has infinitely many solutions in the
interval (9) but no more than one solution outside this interval, i.e. there
are infinitely many branches of bulk MSW but unique branch of surface MSW.

4.6. PHASE VELOCITY VECTOR.

It should be underlined that generally neither bulk nor surface waves are
standing waves with respect to \,$\,z\,$\,-coordinate. The matter is that \,$\,\,\Omega
_{\times }\,$\, and thus \,$\,k_{0}(\omega )\,$\, differ from zero, and actual wave
phase is \,$\,\vartheta (r)=\,$\, \,$\,k_{x}x+\,$\, \,$\,k_{y}y\,$\, \,$\,+k_{0}z\,$\,. In other words
phase velocity vector in MSW is not in-plane oriented but has also non-zero
out-plane component. This fact means that usual attempts to find eigenwaves
assuming the equality \,$\,\,q_{-}=\,$\, \,$\,-q_{+}\,$\, are wrong. According to Eq.3.25,
this equality takes place at special orientation of \,$\,k\,$\, or static
magnetization only. In particular, for \,$\,A=0\,$\,, it is true if \,$\,\nu _{\Vert
}\cos \,\theta \sin \,\theta \,$\, \,$\,=0\,$\,, i.e. if static magnetization \,$\,S_{0}\,$\,
vector is either strictly orthogonal or strictly parallel to film plane or
if \,$\,k\,\,$\, is strictly parallel to \,$\,S_{0}\,$\, projection onto this plane.

4.7. CHARACTERISTIC FREQUENCIES.

Let us discuss characteristic frequencies defined by Eqs.3.26-28. It can be
proved, firstly, that the sum \,$\,\Omega _{zz}+\,$\, \,$\,\Omega _{\nu \nu }\,$\, is always
positive and, secondly, the expression under square root in Eq.3.27 is
always non-negative. Hence, indeed \,$\,\omega _{1}^{2}\,$\, and \,$\,\omega _{2}^{2}\,$\,
are positive values and both the frequencies \,$\,\omega _{1}\,$\, and \,$\,\omega _{2}\,$\,
do exist. Thirdly, the inequalities

\begin{equation}
\omega _{1}^{2}\leq \omega _{u}^{2}\leq \omega _{2}^{2}\, \,\
\end{equation}
take place, that is the uniform precession frequency is either immersed into
bulk waves spectrum or coincides with its edge. This is manifestation of
above mentioned fact that arbitrary internal layer can undergo autonomous
precession.

Next, notice that \,  \,  \,$\,\omega _{3}\geq \omega _{1}\,$\, (see
Eq.3.28). In view of this inequality as combined with (10), the dispersion
equation (7) can not be satisfied at \,  \,  \,$\,\omega <\omega
_{1}\,$\,. At the same time, it is easy to see that each frequency from the
bulk waves interval (9) can be solution to Eq.7 at some appropriate \,$\,|k|\,\,$\,.
Consequently, \,  \,  \,$\,\omega _{1}\,\,$\, is nothing but lower
bound of total MSW spectrum.

It is necessary to remember that all the characteristic frequencies  except \,$\,%
\omega _{u}\,$\, only, are flowing in the sense of their essential dependence on
wave direction, even at \,$\,A=0\,$\,. This is specific anisotropy dictated by flat
geometry of dipole interaction.

From previous reasonings  we must conclude that the surface eigenmodes  if
they exist at all, possess frequencies higher than any bulk wave, with\,$\,%
\;\omega _{2}\,$\, being lower bound of surface wave spectrum. But, evidently,
the Eq.7 does not have roots at\,  \,$\,\omega >\,$\, \,$\,\max (\omega
_{2}\,\omega _{3})\,$\,. Hence, surface waves do exist at those
directions \,$\,\nu =k/|k|\,$\, which satisfy the condition \,$\,\,\omega _{3}^{2}>\,$\, \,$\,%
\omega _{2}^{2}\,$\,. Then Eq.7 has roots at \,$\,\,\kappa (\nu )<\,$\, \,$\,|k|\,$\, \,$\,<\infty \,$\,
with lowest wave number, \,$\,\kappa (\nu )\,$\,, determined by equality \,$\,\,\omega
_{3}^{2}=\omega _{2}^{2}\,$\, :

\begin{equation}
D\kappa (\nu )=(\omega _{2}^{2}-\omega _{u}^{2})/(\omega _{3}^{2}-\omega
_{2}^{2})
\end{equation}

4.8. SURFACE WAVES.

From previous reasonings  we can conclude that the surface eigenmodes  if
they exist at all, must possess eigen-frequencies higher than bulk waves
with\,$\,\;\omega _{2}\,$\, being lower bound of their spectrum. But, evidently, the
Eq.7 does not have roots at\,  \,$\,\omega >\,$\, \,$\,\max (\omega _{2}\,\omega
_{3})\,$\,. Consider more carefully the condition for surface modes
to exist, \,$\,\,\omega _{3}^{2}>\,$\, \,$\,\omega _{2}^{2}\,$\,. With the help of Eq.3.28,
this condition takes the form \

\begin{equation}
(4\pi \sin \,\phi \sin \,\theta )^{2}>\omega _{2}^{2}-\omega _{1}^{2}=4\pi
\sqrt{(\Omega _{zz}+\Omega _{\nu \nu })^{2}-(2\sin \,\phi \sin \,\theta
)^{2}\{(W_{0}+A_{+})^{2}-A_{-}{}^{2}\}}\,
\end{equation}
( \,$\,\phi \,$\, was defined in Eqs.3.17). The Eqs.3.22-23 imply

\begin{equation}
\Omega _{zz}+\Omega _{\nu \nu }=(W_{0}+A_{+})(1+\sin ^{2}\phi \sin
^{2}\theta )+
\end{equation}

\[
+A_{-}[(\sin ^{2}\phi \sin ^{2}\theta +\cos \,2\phi )\cos \,2\psi -\cos
\,\theta \sin \,2\phi \sin \,2\psi ]
\]
Analyzing these formulas  one can see that the requirement (12) can be most
easy satisfied for nearly in-plane static magnetization \,$\,S_{0}\,$\, and then for
waves propagating nearly perpendicular to \,$\,S_{0}\,$\,. Any surface modes with
parallel propagation or in normally magnetized film are clearly forbidden.
From the other hand, taking exactly in-plane \,$\,S_{0}\,$\, and \,$\,\nu _{\bot }^{2}=1\,$\,
, i.e. strictly perpendicular waves  we come from (12) to inequality

\begin{equation}
2\pi >|A_{-}|\,,
\end{equation}
which is practically always true.

Hence, if anisotropy is not extremely strong (in the sense of (14)), then
surface waves definitely exist in some region surrounding the point \,$\,\theta
=\phi \,$\, \,$\,=\pi /2\,$\,. For given \,$\,\theta \,$\, and \,$\,\phi \,$\, in this region, let \,$\,%
\,\omega _{h}\,$\, be their highest (upper) frequency (thus \,$\,\omega _{h}\,$\, is
also upper bound of MSW spectrum). Naturally, \,$\,\omega _{h}\,$\, is achieved at \,$\,%
|k|\rightarrow \infty \,$\,  when \,$\,\cot \,[\Lambda |k|D]\,$\, \,$\,\rightarrow
1\,$\,, therefore it follows from Eq.7 that

\begin{equation}
\omega _{h}^{2}=\frac{\omega _{3}^{4}-\omega _{1}^{2}\omega _{2}^{2}}{(4\pi
\sin \,\phi \sin \,\theta )^{2}}=\left( \frac{\Omega _{zz}+\Omega _{\nu \nu }%
}{2\sin \,\phi \sin \,\theta }+2\pi \sin \,\phi \sin \,\theta \right) ^{2}
\end{equation}
Particularly, for strictly in-plane magnetization and perpendicular
propagation, \,$\,\,k\bot S_{0}\,$\,, this expression reduces to

\begin{equation}
\omega _{h}(\theta =\pi /2,\phi =\pi /2)=W_{0}+A_{+}+2\pi
\end{equation}
We can expect this is maximum \,$\,\,\omega _{h}\,$\,.

4.9. ISOTROPIC CASE.

All the algebra becomes much more visual if anisotropy contribution
disappears. This does not necessarily mean that anisotropy is absent at all.
For example, in the easy axis or easy plane case, anisotropy energy (2.2)
and matrix \,$\,\widehat{A}\,$\, (see Sec.2, Eq.2.11) take the form

\[
A(S)=A_{0}\left\langle \overline{u}\,S\right\rangle ^{2}/2\,,\,  \
\widehat{A}=A_{0}\,\overline{u}\otimes \overline{u}\,\,,
\]
where \,$\,\overline{u}\,$\, is unit vector showing easy (or heavy) axis. Hence, at
special static magnetization, when \,$\,S_{0}\parallel \,\overline{u}\,$\,, the
projected anisotropy matrix exactly turns into zero: \,$\,\widehat{\Pi }\widehat{%
A}\widehat{\Pi }=0\,$\,.

In absence of anisotropy contribution, the inequality (10) reduces to

\begin{equation}
\omega _{1}^{2}=W_{0}^{2}+4\pi W_{0}\nu _{\bot }^{2}\sin ^{2}\theta \,\leq
\,\omega _{u}^{2}=W_{0}^{2}+4\pi W_{0}\sin ^{2}\theta \,\leq \,\omega
_{2}^{2}=W_{0}^{2}+4\pi W_{0}\,\,,\, (A=0)
\end{equation}
We see that under in-plane magnetization, i.e. at \,$\,\theta =\pi /2\, \,$\,, (i) \,$\,%
\,\omega _{2}=\omega _{u}\,$\,, that is all the bulk wave spectrum lies under
uniform precession frequency,\,  and (ii) there are no bulk waves
propagating perpendicular to static magnetization, because the equality \,$\,%
\omega _{1}=\omega _{2}\,\,$\, takes place at \,$\,\,\nu _{\bot }^{2}\equiv \,$\, \,$\,%
\sin ^{2}\phi \,$\, \,$\,=1\,$\,.

The condition (12) determining the surface modes region now reads

\begin{equation}
\sin ^{2}\phi \sin ^{2}\theta >\sin ^{2}\theta _{0}\equiv \frac{W_{0}}{%
W_{0}+4\pi }\,\,,\, \text{\,  }\,  (A=0)\,,
\end{equation}
while their upper frequency is presented by

\[
\omega _{h}^{2}=W_{0}(W_{0}+4\pi )\frac{1}{4}\left( \frac{\sin \,\theta _{0}%
}{\sin \,\phi \sin \,\theta }+\frac{\sin \,\phi \sin \,\theta }{\sin
\,\theta _{0}}\right) ^{2}\,,\,  (A=0)
\]
In accordance with (16), its absolute maximum is \,$\,\,\max \,\omega _{h}=\,$\, \,$\,%
W_{0}+2\pi \,$\,.

The lowest wave number of surface modes is achieved at \,$\,k\,$\, strictly
perpendicular to \,$\,z\,$\,-\,$\,S_{0}\,$\,-plane. From Eq.11 we obtain

\begin{equation}
\min \,D\kappa (\nu )=D\kappa (\nu \bot S_{0})=\frac{2W_{0}}{4\pi \tan
^{2}\theta -W_{0}}
\end{equation}
Naturally, it turns into infinity at \,$\,\theta =\theta _{0}\,$\, when surface
modes disappear. For in-plane magnetization it turns into zero, and then
surface waves occupy all the sector

\begin{equation}
\left| \frac{\nu _{\Vert }}{\nu _{\bot }}\right| =\left| \frac{k_{\Vert }}{%
k_{\bot }}\right| <\sqrt{\frac{4\pi }{W_{0}}}
\end{equation}

It should be added that under in-plane external field \,$\,\theta =\pi /2\,\,$\,,
and the static internal field, \,$\,W_{0}\,$\,, trivially reduces to the external
one, \,$\,W_{0}=H_{0}\,$\,, because of zero demagnetization.

4.10. DAMON-ESHBACH WAVES.

As was mentioned, at \,$\,A=0\,$\,, in-plane magnetization ( \,$\,\theta
=\pi /2\,\,$\,) and \,$\,\,$\, \,$\,\sin ^{2}\phi \,$\, \,$\,=1\,$\, all three frequencies present in
(10) coincide one with another. Hence, in this specific case
identically\,  \,$\,\Lambda (\omega )=1\,$\,, and with accounting for (17)
the Eq.7 becomes linear equation:

\[
\coth \,(|k|D)\, =(\omega _{3}^{2}-\omega ^{2})/(\omega ^{2}-\omega
_{u}^{2})\,,\,   \omega _{3}^{2}=\omega _{u}^{2}+8\pi ^{2}\,\
,\, \omega _{u}^{2}=H_{0}(H_{0}+4\pi )
\]
Its solution is given by the classical Damon-Eshbach dispersion law (3.38).
Otherwise, unfortunately, Eq.7 can not be solved in such an evident form.
But in wide sense all the surface modes can be called Damon-Eshbach waves.

\begin{figure}
\includegraphics{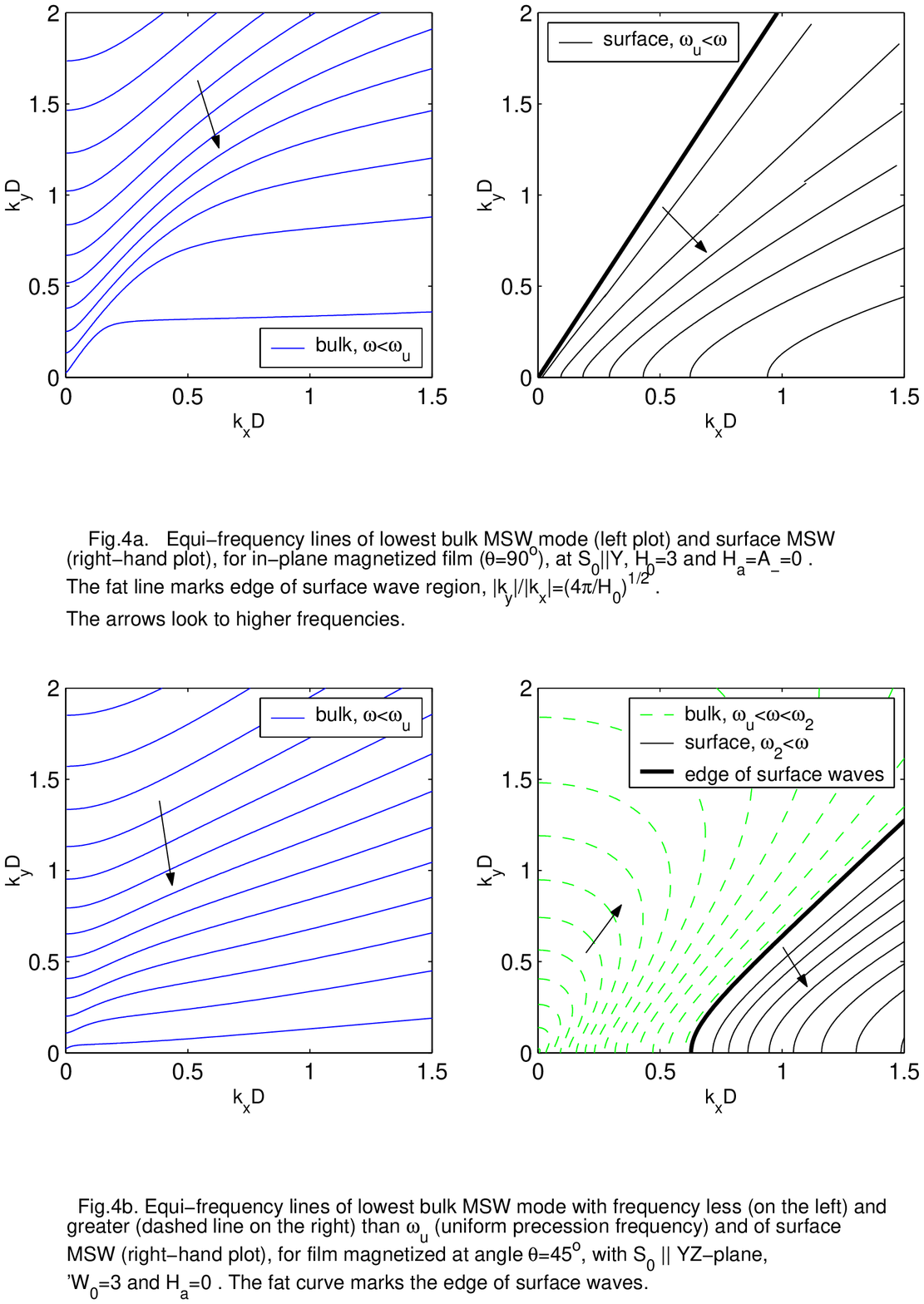}
\end{figure}

4.11. MSW CLASSIFICATION.

In general, dispersion law must be obtained numerically. But also we may
treat Eq.7 be evident expression for \,$\,|k|\,$\, as a function of wave direction
and wave frequency:

\begin{equation}
|k|D\, =%
\mathop{\rm Re}%
\frac{|\omega ^{2}-\omega _{u}^{2}|}{\sqrt{(\omega ^{2}-\omega
_{1}^{2})(\omega _{2}^{2}-\omega ^{2})}}\left\{ \pi N+\arctan \frac{\text{%
sign\,  }(\omega ^{2}-\omega _{u}^{2})\sqrt{(\omega ^{2}-\omega
_{1}^{2})(\omega _{2}^{2}-\omega ^{2})}}{\omega _{3}^{2}-\omega ^{2}}%
\right\} \,,
\end{equation}
where \,$\,N\,$\, is non-negative integer, the square root should be chosen in upper
half-plane, while \,$\,\arctan \,$\, in right-hand half-plane, and the requirement \,$\,%
|k|D\geq 0\,$\, serves as selection rule for permissible frequencies. Under
these conditions  \,$\,N\,$\, plays no role outside of interval (9) while inside it \,$\,%
N\geq 0\,$\, enumerates branches of bulk waves with different normal
wavenumbers  \,$\,|k|D|\Lambda (\omega )|\,$\,. In fact at any \,$\,N\,\,$\, two modes can
be distinguished, with \,$\,\omega <\omega _{u}\,$\, and \,$\,\omega >\omega _{u}\,$\,, to
be enumerated as \,$\,N+\,$\, and \,$\,N-\,$\,, respectively.

4.12. MAIN MODES.

Fig.4a shows equal-frequency lines calculated from Eq.22 for the lowest bulk
mode \,$\,0-\,\,$\,and surface mode under in-plane magnetization and in absence of
anisotropy (for details  see captures). That are just the two sorts of MSW
discussed in Sec.3 as contributors to impedances at Fig.2a-2b and 1a-1d,
respectively.

Wonderfully, their frequency spectra, although lying on opposite hands from
\,$\,\omega _{u}\,$\,, at \,$\, |k|D\rightarrow 0\,$\, are sewed together in the
directions (21), i.e. along the edge of surface wave region. Therefore, with
respect to sufficiently long waves both the sorts can be effectively unified
into single main mode. Corresponding expansion of Eq.7 or Eq.22 gives its
dispersion law as follows:

\begin{equation}
\omega =\omega _{0}(kD)\,,\,   \omega _{0}(kD)\approx \omega _{u}+%
\frac{\omega _{3}^{2}-\omega _{u}^{2}}{2\omega _{u}}|k|D=\omega _{u}+\frac{%
\pi D(4\pi k_{x}^{2}-H_{0}k_{y}^{2})}{|k|\sqrt{H_{0}(H_{0}+4\pi )}}
\end{equation}
This equation extends the Damon-Eshbach formula (3.38) to arbitrary
propagation angles although at small wave numbers. The inequality\,  \,$\,%
\,\,|k|D\lesssim 0.2\,$\, is quite sufficient to apply Eq.23.

Fig.4b shows what does occur if static magnetization is put out from the
film plane. We see that now main bulk mode \,$\,0-\,\,$\,becomes strongly separated
from surface wave but instead the latter well merges with mode \,$\,0+\,$\,.

\begin{figure}
\includegraphics{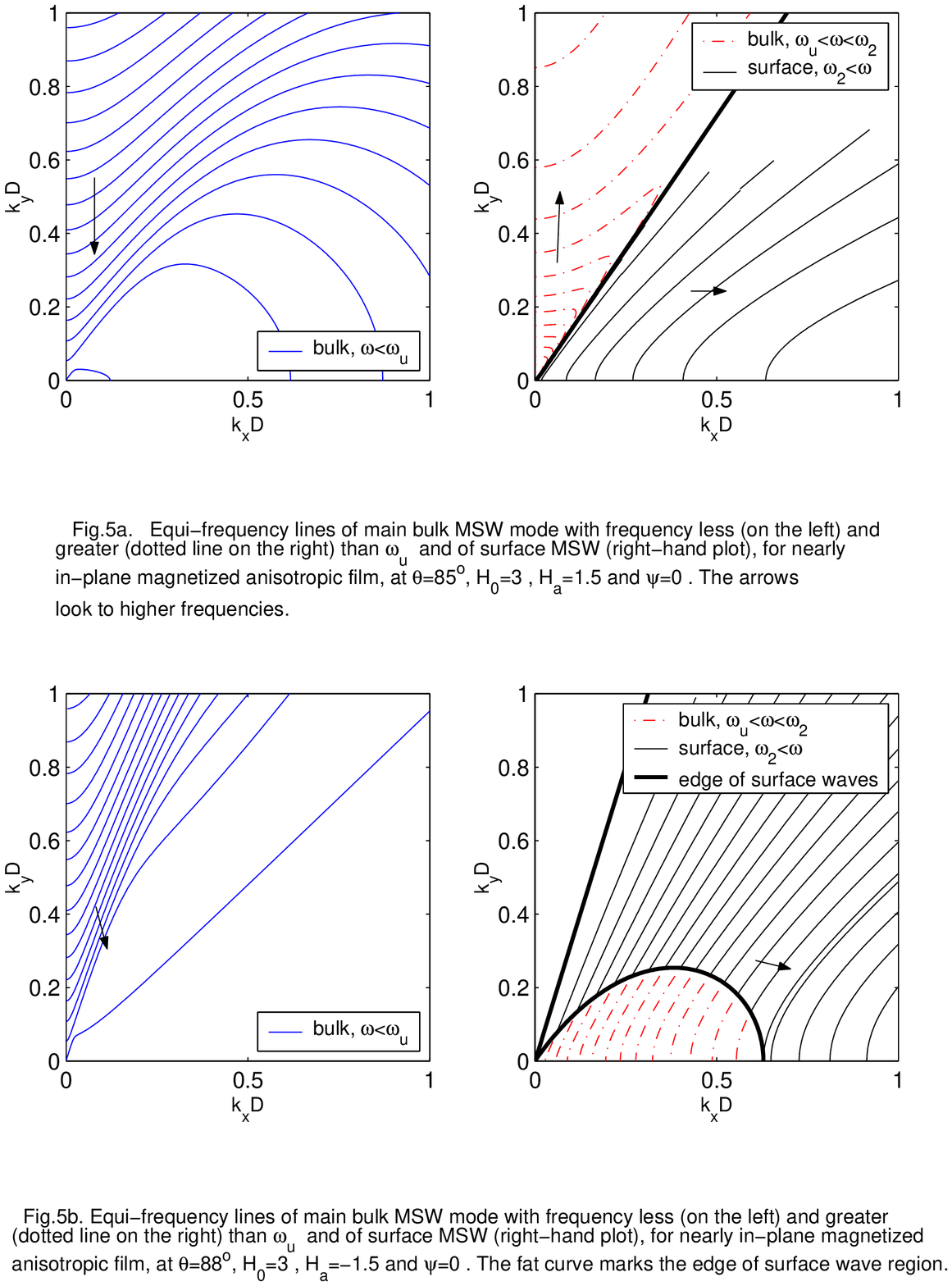}
\end{figure}

4.13. EFFECTS OF ANISOTROPY.

To feel principal influence by anisotropy, let us confine ourselves by
special case \,$\,\psi =0\,$\,, when one of main axes of \,$\,\widehat{\Pi }\widehat{A}%
\widehat{\Pi }\,$\,, namely \,$\,\overline{A}_{1}\,$\,, lies in \,$\,z\,$\,-\,$\,S_{0}\,$\,-plane while
\,$\,\overline{A}_{2}\,$\, in \,$\,x\,$\,-\,$\,y\,$\,-plane. Then

\begin{equation}
\omega _{u}^{2}=W_{2}(W_{1}+4\pi \sin ^{2}\theta )\,,\,  (\psi =0)\,,
\end{equation}

\begin{equation}
\omega _{1,2}^{2}=W_{1}W_{2}+2\pi \lbrack W_{1}\sin ^{2}\phi +W_{2}(1-\cos
^{2}\theta \sin ^{2}\phi )]\mp 2\pi \lbrack (W_{1}\sin ^{2}\phi -W_{2}\sin
^{2}\theta )^{2}+
\end{equation}

\[
+W_{2}^{2}\cos ^{2}\theta \cos ^{2}\phi (1+\sin ^{2}\theta -\cos ^{2}\theta
\sin ^{2}\phi )+2W_{1}W_{2}\cos ^{2}\theta \sin ^{2}\phi \cos ^{2}\phi
]^{1/2}\,,
\]
where \,$\,W_{1,2}=W_{0}+A_{1,2}=\,$\, \,$\,W_{0}+A_{+}\mp A_{-}\,$\,. Instead of (20) one
can obtain

\begin{equation}
\min \,D\kappa (\nu )=D\kappa (\nu \bot S_{0})=\frac{X+|X|}{4\pi \sin
^{2}\theta -|X|}\,,\,  X\equiv W_{1}-W_{2}\sin ^{2}\theta \,,\,\
(\psi =0)
\end{equation}
Clearly, the role of \,$\,A_{+}\,$\, is merely shift of the internal field, and
non-trivial effects may come from the effective anisotropy field \,$\,%
\,H_{a}\equiv A_{-}\,$\, only.

For example, consider nearly in-plane magnetized film (\,$\,\theta =85^{o}\,$\,)
with such an anisotropy. From Eq.26 it is seen that at\,$\, H_{a}>0\,\,$\, (\,$\,%
W_{1}<\,$\, \,$\,W_{2}\,$\,) and \,$\,\theta \,$\, close to \,$\,90^{o}\,$\, the beginning of the
surface wave sector remains staying at \,$\,k=0\,$\,. That is illustrated by
Fig.5a. It shows also that interface between surface mode and \,$\,0+\,$\, bulk mode
is rather sharp, so that the \,$\,0-\,$\, mode (left plot) seems be better
continuation of the surface sector. Indeed, at \,$\,\theta \rightarrow \,$\, \,$\,90^{o}\,$\,
the picture becomes very similar to Fig.4a. Hence, positive anisotropy
field, \,$\, H_{a}>0\,\,$\,, (i) produces no qualitative change in relation between
the two sorts of waves. However, it (ii) allows \,$\,0-\,$\, mode to propagate in \,$\,x\,$\,%
-direction, (iii) in accordance with Eq.24, it narrows down the surface wave
sector, and (iv) rises \,$\,\omega _{u}\,$\, and thus MSW frequencies (as reflected
by Fig.1a too).

In contrast, Fig.5b demonstrates that negative anisotropy, \,$\, H_{a}<0\,\,$\, (\,$\,%
W_{1}>\,$\, \,$\,W_{2}\,$\,), (i) causes essential change in relative disposition of
bulk and surface modes on \,$\,k\,$\,-plane. Namely, the \,$\,0+\,\,$\, mode becomes
captured in the bubble immersed into the surface sector (the right-hand edge
of this bubble is determined by Eq.26). The matter is that at \,$\,\theta
\approx 90^{o}\,$\,, according to Eqs.24 and 25,

\[
\omega _{u}^{2}=\omega _{1}^{2}\,  \,\text{at\,   }\sin ^{2}\phi
>W_{2}/W_{1}\,,\,   \omega _{u}^{2}=\omega _{2}^{2}\,  \,\text{%
at\,   }\sin ^{2}\phi <W_{2}/W_{1}
\]
This results also in (ii) precise angular separation of \,$\,0-\,\,$\,(left plot)
and \,$\,0+\,\,$\, modes  which facilitates to treat, at \,$\,|k|D<<1\,$\,, all three modes
as single anisotropic mode. Besides  negative anisotropy (iii) obviously
expands the surface sector and (iv) decreases \,$\,\omega _{u}\,$\, and MSW
frequencies.

4.14. LONG-WAVE ASYMPTOTICS AND EXCHANGE CONTRIBUTION.

In general (at \,$\,\sin ^{2}\theta <1\,$\, ), the dispersion law for long MSW, as
compared with film thickness  can approximated by linear function of \,$\,|k|D\,$\,
, like in Eq.23. The Eq.7 yields

\begin{equation}
\omega _{N\pm }^{2}-\omega _{u}^{2}\approx \frac{|k|D(\omega _{3}^{2}-\omega
_{u}^{2})X}{\pi N+\arctan \,X+\pi (1-\text{sign\,  }X)/2}\,,\,\,\
X\equiv \pm \frac{\sqrt{(\omega _{u}^{2}-\omega _{1}^{2})(\omega
_{2}^{2}-\omega _{u}^{2})}}{\omega _{3}^{2}-\omega _{u}^{2}}\,
\end{equation}
Hence, MSW modes with greater \,$\,N\,$\, are more strongly pressed to the uniform
precession frequency. This paradoxical fact is the consequence of above
discussed singularity of dipole interaction.

Under strictly in-plane magnetization ( \,$\,\sin ^{2}\theta =1\,$\, ), at \,$\,N=0\,$\,
this expression reduces to Eq.23 which unify main modes (may be with the
surface one instead of \,$\,0+\,\,$\,). But for \,$\,N>0\,$\, dispersion becomes quadratic
leading to even stronger frequency compression:

\begin{equation}
\omega _{N\pm }^{2}-\omega _{u}^{2}\approx \pm (|k|D)^{2}(\omega
_{u}^{2}-\omega _{1}^{2})/(\pi N)^{2}\,,\,  (N>0)
\end{equation}
However, at sufficiently large \,$\,N\,$\, the exchange interaction enters the game
and increases \,$\,\omega _{N\pm }\,$\,. At least under in-plane magnetization, the
exchange contribution can be described by the replacement

\begin{equation}
W_{0}\Rightarrow W_{0}+r_{0}^{2}(|k|^{2}+q_{N\pm }^{2})\,\approx
W_{0}+r_{0}^{2}(\pi N/D )^{2}\,,
\end{equation}
with \,$\,q_{N\pm }\,$\, being out-plane wave numbers  \,$\,q_{\pm }\,$\, (see Eq.8), for \,$\,N\,$\,%
-th modes  in exact analogy with how exchange interaction contributes to
frequency of boundless wave (Sec.2.11). In the latter equality, we took into
account that for higher order long-wave modes \,$\,r_{0}|k|\,$\, is negligibly small
while their normal wave numbers \,$\,q_{\pm }=\,$\, \,$\,\pm |k|\Lambda (\omega )\,$\, are
close to \,$\,\pi N/D\,$\,.



\section{Some aspects of non-linear phenomena and chaos}

5.1. NONLINEAR PROCESSES.

The unique peculiarity of spin waves (SW) and especially magnetostatic waves
(MSW) in ferrimagnets (e.g. YIG) is that their relaxation rate, \,$\,\Gamma \sim
\,$\, \,$\,5\,\,$\,[\,$\,\mu \,$\,s\,$\,^{-1}\,$\,] or even less  is very small as compared with other
wave excitations in solids in the same (microwave) frequency region. So low
decay ensures effective generation and nonlinear transformations of MSW at
small pumping power [1-5].

At the same time, usually swing of spin precession remains far from spin
flipping, i.e. \,$\,|S_{\bot }|^{2}<<1\,$\, in Eq.2.19, therefore three-wave and
four-wave processes only are of great importance. In the firsts  either (P)
some already excited mode with frequency \,$\,\omega _{0}\,$\, serves as parametric
pump for two other modes whose frequencies satisfy the condition \,$\,\omega
_{1}+\omega _{2}\,$\, \,$\,=\omega _{0}\,$\, or, in opposite, (G) two modes mix up one
another being the source for \,$\,\omega _{0}\,$\, mode. Here the bracketed letters
G and P, abbreviate generation and parametric excitation. Among fourth-order
processes  most important one is the combined G-P-process satisfying \,$\,%
\,\omega _{3}+\omega _{4}\,$\, \,$\,=\omega _{1}+\omega _{2}\,$\,.

If accounting for these processes only, the Eq.2.19 transforms into the
approximate equation,
\begin{equation}
\frac{dS_{\bot }}{dt}=[S_{0}\,(W_{0}+\widehat{L})\,S_{\bot }-h]-\gamma \widehat{%
\Pi }(W_{0}+\widehat{L})\,S_{\bot }-
\end{equation}

\[
-\left\langle S_{0}\,\widehat{L}\,S_{\bot }-h\right\rangle [S_{0}\,S_{\bot
}]-[S_{0}\,\widehat{L}(|S_{\bot }|^{2}\,S_{0})]/2-
\]

\[
-|S_{\bot }|^{2}[S_{0}\,\widehat{L}\,S_{\bot }]/2+\left\langle S_{0}\,\widehat{L}%
(|S_{\bot }|^{2}\,S_{0})\right\rangle [S_{0}\,S_{\bot }]/2\,,\, \,  \
\widehat{L}=\widehat{A}-r_{0}^{2}\nabla ^{2}+\widehat{G}\,,
\]
where three rows contain linear, quadratic and cubic terms  respectively.
Higher-order terms and all the nonlinear contributions to friction (as well
as to anisotropy, see Sec.2.5) are neglected, and most important entries of \,$\,%
h(r,t)\,$\, are kept only.

Clearly, the external field, \,$\,h(r,t)\,$\,, also can act as either additive
source (G-process) or parametric pump (P-process). The first variant is more
effective if realizes by way of ferromagnetic resonance (FMR). In best real
YIG samples  the power consumption of order of tens microwatt may be
sufficient to initiate nonlinear processes [2]. At greater pump, one can
observe rich variety of nonlinear phenomena including formation of envelope
solitons [2,3-9], parametric amplification [10-12], magnetization reversal
[13], self-focusing of MW beams[14], generation of harmonics  subharmonics
and ultra-short pulses [15], non-linear short electromagnetic waves [16]
(the alternate to MSW high-frequency branch of  mutual magnetization and
EM-field hybridization).

But most interesting phenomenon is magnetic chaos (chaotic oscillations of
magnetization pattern) produced if external pump exceeds certain critical
level [2,17-19].

5.2. NONLINEAR WAVES.

In special class of nonlinear phenomena qualified as weakly nonlinear
magnetic waves  a narrow region of total MSW frequency band (all the more of
whole MW spectrum) is involved only and, hence, third-order processes
(quadratic terms in second row of Eq.1) are not at business. In sufficiently
long waves  their dispersion (spatial derivatives) also is weak and
therefore naturally separates from nonlinearity, so that approximate wave
equation turns into sum of spatially non-local (differential) linear terms
and local nonlinear (cubic) terms [3].

If speak about films  the long magnetostatic nonlinear waves are of special
interest composed by the main branch of linear MSW (most homogeneous with
respect to normal \,$\,z\,$\,-coordinate). In this case the exchange part of
operator \,$\,\widehat{L}\,\,$\, can be neglected, and dispersion is completely
determined by dipole interaction (see Sec.4). But since the singular part of
dipole interaction (Sec.4.1) is factually local, its product with cubic
nolinearity should be kept. Then the Eq.1 (as combined with Eq.4.1) reduces
to

\begin{equation}
\frac{dS_{\bot }}{dt}=[S_{0}\,(W_{0}+\widehat{A}+\widehat{G})\,S_{\bot
}-h]-\gamma \widehat{\Pi }(W_{0}+\widehat{A}+4\pi \overline{z}\otimes
\overline{z})\,S_{\bot }+
\end{equation}

\[
+\frac{1}{2}|S_{\bot }|^{2}[S_{0}\,(A_{\parallel }+4\pi S_{0z}^{2}-\widehat{A}%
-4\pi \overline{z}\otimes \overline{z})\,S_{\bot }]\,,\,  \,A_{\parallel
}\equiv \left\langle S_{0}\,\widehat{A}\,S_{0}\right\rangle \,,
\]
where characteristic frequency of \,$\,\,h\,$\, is supposed the same as carrying
frequency of \,$\,S_{\bot }\,\,$\,. Of course, still this is formal storage only for
more correct equation which must be free of third-order harmonics and
concern \,$\,S_{\bot }\,$\,'s envelope. Such the equation can be deduced, as usually
[20], from variational formulation of Eq.1, or by means of time averaging
over the carrier period.

5.3. NONLINEAR WAVE EQUATION.

In accordance with Sec.2.9 and Sec.4.3, the main (as well as any other)
branch of eigenwave modes looks as

\[
S_{\bot }\propto V_{k}(r)\exp \{-i\omega _{0}(kD)t\}\,,\, \
V_{k}(r)=\{a_{k}(z)+ib_{k}(z)\}\exp \{i\left\langle k,\rho \right\rangle \}\,
\]
Here its dispersion law is written in the form \,$\,\,\omega =\,$\, \,$\,\omega _{0}(kD)\,$\,
, \,$\,\,k=\,$\, \,$\,\{k_{x}\,k_{y}\}\,$\, is in-plane wave vector, \,$\,\rho \equiv \{x,y\}\,$\,%
\,, and \,$\,a_{k}(z)\,$\, and \,$\,b_{k}(z)\,\,$\, are mutually orthogonal
real-valued vectors. Arbitrary non-autonomous (externally influenced) wave
composed by these modes can be expanded into Fourier integral

\begin{equation}
S_{\bot }=%
\mathop{\rm Re}%
\,\int \{a_{k}(z)+ib_{k}(z)\}e^{i\left\langle k, %
\rho \right\rangle }C(k) %
\widetilde{\Psi }(k,t)\, dk\,,
\end{equation}
where function \,$\,\widetilde{\Psi }(k,t)\,$\, contains one-signed (e.g. positive)
frequencies only, that is represents an analytical signal. The \,$\,C(k)\,$\, in (3)
being real positive factor serves for suitable normalization of the
eigenmodes. If it is fixed then, instead of (3), one can equivalently
consider the ``wave function''

\begin{equation}
\Psi =\Psi (x,y,t)\equiv e^{i\left\langle k,\rho \right\rangle }%
\widetilde{\Psi }(k,t)dk
\end{equation}
Correspondingly to (3), it useful to introduce analytical signal, \,$\,%
\widetilde{h}\,$\,, for the external pump too:

\[
h(r,t)=%
\mathop{\rm Re}%
\,\widetilde{h}(r,t)\,,\, \,  \int e^{i\omega t}\widetilde{h} %
(r,t)\, dt\equiv 0\,\, \,  \text{at\,  }\, \omega \leq 0
\]

For the wave function, the Eq.2 implies the equation as follows (we omit its
derivation):

\begin{equation}
\frac{\partial \Psi }{\partial t}+i\omega _{0}(-iD\nabla )\Psi =-i\varkappa
|\Psi |^{2}\Psi -\Gamma \Psi +\eta \,
\end{equation}
Here \,$\,\nabla =\{\partial /\partial x,\partial /\partial y\}\,\,$\,and operator \,$\,%
\omega _{0}(-iD\nabla )\,$\, (formally differential) is determined by the
dispersion law. Let \,$\,a,\,b\,$\, and \,$\,\alpha,\,\beta \,$\, be the pair of
eigenvectors and related eigenvalues of the uniform precession operator, \,$\,%
\widetilde{W}\equiv \,$\, \,$\,\widehat{\Pi }(W_{0}+\widehat{A}+4\pi \,\overline{z%
}\otimes \overline{z})\widehat{\Pi }\,\,$\,, considered in Sec.4.3, and \,$\,%
\,p\,\,\,$\,eccentricity of uniform precession. Besides  for any two vectors \,$\,u
\,$\, and \,$\,\,v\,$\,, let \,$\,\,u_{v}\,\,$\,means \,$\,u\,$\,'s projection onto \,$\,v\,\,$\,,
i.e. \,$\,u_{v}\equiv \left\langle v,u\right\rangle /|v|\,$\,. Then the parameters
of Eq.5, friction coefficient \,$\,\Gamma \,$\,, nonlinearity scale\,  \,$\,%
\varkappa \,$\, and pump, \,$\,\eta \,$\,, read

\begin{equation}
\Gamma =(\alpha +\beta )\gamma /2\,,
\end{equation}

\begin{equation}
\eta =\frac{1}{2D}\int \left\{ \sqrt{p}\widetilde{h}_{a}- %
i\widetilde{h}_{b}/ %
\sqrt{p}\right\} \, dz\,\,,\,\, \, \,  %
p =\sqrt{\frac{\alpha }{\beta }}\,,
\end{equation}

\begin{equation}
\varkappa =\frac{1}{8}\left\{ (W_{0}+A_{\parallel }+4\pi S_{0z}^{2})\left[
\frac{3}{2}\left( \frac{\alpha }{\beta }+\frac{\beta }{\alpha }\right) +1%
\right] -2(\alpha +\beta )\right\} \,,
\end{equation}
while approximate connection between the wave function and magnetization is
established by

\begin{equation}
\Psi =\sqrt{p}\,S_{\bot a}-iS_{\bot b}/\sqrt{p}\,,\,   \,  |\Psi
|^{2}\approx \left\| S_{\bot a}\right\| \left\| S_{\bot b}\right\|
\end{equation}
Here \,$\,\left\|...\right\| \,$\, denotes envelope (amplitude) of an oscillating
variable.

In particular case of tangential magnetization and not strong anisotropy,
vector \,$\,a\,\,$\,is nearly parallel to normal \,$\,z\,$\,-axis  vector \,$\,b\,\,$\, lies in
the film plane, and formulas (6-8) are simplified to

\begin{equation}
\Gamma \approx (H_{0}+2\pi )\gamma \,,\,  p\approx \sqrt{1+4\pi /H_{0}}%
\,,\,   \varkappa \approx -\frac{\pi (H_{0}+\pi )}{H_{0}+4\pi }
\end{equation}

5.4. NONLINEAR SHR\"{O}DINGER EQUATION.

For waves and wave packets formed by a narrow set of in-plane wavevectors
concentrated about some \,$\,k_{0}\,,\,$\,the Eq.5 reduces to the nonlinear Shr\"{o}%
dinger equation (NLS), \

\begin{equation}
\frac{\partial \psi }{\partial t}+\left\langle v_{g}\,\nabla \right\rangle
\psi =i\left\langle \nabla,\widehat{D}\nabla \right\rangle \psi -i\varkappa
|\psi |^{2}\psi -\Gamma \psi +\widetilde{\eta }\,,
\end{equation}

\[
\psi \equiv \exp \{i\omega _{0}(Dk_{0})t-i\left\langle k_{0}\,\rho
\right\rangle \}\Psi \,,\,   \widetilde{\eta }\equiv \exp \{i\omega
_{0}(Dk_{0})t-i\left\langle k_{0}\,\rho \right\rangle \}\eta \,,
\]

\begin{equation}
v_{g}=\frac{\partial \omega _{0}(Dk_{0})}{\partial k_{0}}\,,\,  \widehat{%
D}_{mn}=\frac{1}{2}\frac{\partial ^{2}\omega _{0}(Dk_{0})}{\partial
k_{0m}\partial k_{0n}}\,
\end{equation}
Here \,$\,v_{g}\,$\, and \,$\,\widehat{D}\,\,$\, are group velocity vector and diffusivity
tensor, respectively, and \,$\,\psi \,$\, plays the role of envelope of wave
function \,$\,\Psi \,$\,.

Evident analytical expressions for the latter quantities can be obtained in
a few special cases only, particularly, for Damon-Eshbach waves (see Sec.3
and Sec.4) in exactly in-plane magnetized film with zero (or weak)
anisotropy. In this case, if magnetizing field \,$\,H_{0}\,$\, is oriented along \,$\,y\,$\,%
-axis then for surface waves nearly parallel to \,$\,x\,$\,-axis (\,$\,k_{y}^{2}\ll
|k|^{2}\,$\,) \,  the Eq.4.7 yields (in the dimensionless time units):

\begin{equation}
\omega _{0}^{2}(Dk)\approx \omega _{u}^{2}+\pi \{1-\exp (-2D|k|)\}(4\pi
k_{x}^{2}-H_{0}k_{y}^{2})/|k|^{2}
\end{equation}
Hence, wave packet running along \,$\,x\,$\,-axis  with \,$\,k_{0y}=0\,$\,, has group
velocity and diffusivity as follow:

\begin{equation}
v_{gy}=0\,\,,\,   v_{gx}=4\pi ^{2}D\exp (-2D|k_{0}|)\text{sign}%
(k_{0})/\omega _{0}\,,\,   \omega _{0}\equiv \omega _{DE}(D|k_{0}|)\,,
\end{equation}
with function \,$\,\,\omega _{DE}(D|k|)\,\,$\,given by Eq.3.38, and

\begin{equation}
\widehat{D}_{xx}=-4\pi ^{2}D^{2}\{(H_{0}+2\pi )^{2}-2\pi ^{2}\exp
(-2D|k_{0}|)\}\exp (-2D|k_{0}|)/\omega _{0}^{3}\,,
\end{equation}

\begin{equation}
\widehat{D}_{xy}=0\,\,,\,  \widehat{D}_{yy}=-\pi (H_{0}+4\pi )\{1-\exp
(-2D|k_{0}|)\}/2|k_{0}|^{2}\,
\end{equation}

In contrary to this specific case, generally propagation direction of
envelope of the wave packet differs from its carrier wave direction, \,$\,k_{0}\,$\,%
\,  \,. Clearly, the group velocity is perpendicular
to equi-frequency curves shown at Fig.4a-b and Fig.5a-b (see Sec.4).
These figures (as well as formulas of Sec.4) show that wave packets
which are formed by surface MSW and have non-zero \,$\,k_{0y}\,$\, comparable
with \,$\,k_{0x}\,\,$\, must prefer directions characterized by

\begin{equation}
v_{gy}\approx \pm v_{gx}\sqrt{H_{0}/4\pi }
\end{equation}

If the carrier wave is not long, that is the value \,$\,D|k_{0}|\,\,$\, is
comparable with unit, then the main axes of polarization ellipse, \,$\,a,\,b\,$\,,
its eccentricity, \,$\,p\,\,$\,, and the eigenvalues \,$\,\alpha,\,\beta \,$\, in
Eqs.6-10 should be calculated just for the\,  \,$\,k_{0}\,$\, mode (instead
of uniform one), i.e. mentioned as \,$\,\,a_{k_{0}}\,\,$\,, \,$\,b_{k_{0}}\,\,$\,, and so
on, in the sense of Sec.2.9-10.

5.5. NON-ISOCHRONITY AND INSTABILITY OF MAGNETIC WAVES.

Consider autonomous waves  i.e. in absence of pump and dissipation. Cubic
nonlinear terms in Eq.5 and Eq.11 involve fundamental non-isochronity
property of nonlinear MW: their frequencies depend on their amplitudes.
Indeed, for a plane autonomous wave with amplitude \,$\,A\,$\, the Eq.5 gives

\begin{equation}
\Psi =A\exp \{-i[\omega _{0}(Dk)+\varkappa A^{2}]t+i\left\langle k,\rho
\right\rangle \}
\end{equation}
According to Eq.10, in tangentially magnetized film intensification of wave
leads to lowering its frequency.

What does occur if the amplitude is not uniform but slightly spatially
modulated? As in general [20], result depends on concurrence between
nonlinearity and dispersion which in oure case is described by diffusional
term in Eq.11. To see the result, let us search for evolution of the wave
envelope in the form

\begin{equation}
\psi (\rho,t)=[A+\chi (\rho -v_{g}t,t)]\exp (-i\varkappa A^{2}t)\,,\,\,\
\, \,\chi =\chi _{1}+i\chi _{2}\,,\,   \,A=const\,,
\end{equation}
with \,$\,\chi \,$\,being (infinitely) small non-uniform perturbation. It is easy
to derive from Eq.11 the linearized equations for \,$\,\chi _{1}(\rho,t)\,\,$\,\
and \,$\,\chi _{2}(\rho,t)\,\,$\, as follows

\begin{equation}
\frac{\partial }{\partial t}\left(
\begin{array}{c}
\chi _{1}\, \\
\chi _{2}
\end{array}
\right) =\left(
\begin{array}{cc}
0 & -\left\langle \nabla,\widehat{D}\nabla \right\rangle  \\
\left\langle \nabla,\widehat{D}\nabla \right\rangle -2\varkappa A^{2} & 0
\end{array}
\right) \left(
\begin{array}{c}
\chi _{1}\, \\
\chi _{2}
\end{array}
\right)
\end{equation}
Let initially the amplitude was periodically modulated with some wave vector
\,$\,q\,$\,, for instance, \,$\,\chi _{1}(\rho,0)\propto \,$\, \,$\,\cos \,\left\langle k,\rho
\right\rangle \,$\,,\,  \,$\,\chi _{2}(\rho,0)=0\,\,$\,. Then solution to
(20) consists of two definitely weighted exponents  \,$\,\exp (\pm \lambda t)\,\,$\,%
\,, where \,$\,\pm \lambda \,$\,  are eigenvalues of the
right-hand matrix operator,

\begin{equation}
\lambda ^{2}=-\left\langle q,\widehat{D}q\right\rangle \left( \left\langle q,%
\widehat{D}q\right\rangle +2\varkappa A^{2}\right)
\end{equation}

Here \,$\,\left\langle q,\widehat{D}q\right\rangle \,$\,represents deviation of
wave frequency coming from the dispersion. If it is of the same sign as the
deviation due to non-isochronity then \,$\,\lambda \,$\,has imaginary value.
Hence, in this case initial non-uniformity results in small amplitude and
phase oscillations  which in reality decay due to dissipation. But if

\begin{equation}
\varkappa \left\langle q,\widehat{D}q\right\rangle <0\, \, \text{%
and\,  }\, 2|\varkappa |A^{2}>\left| \left\langle q,\widehat{D}%
q\right\rangle \right| +\Gamma ^{2}\left| \left\langle q,\widehat{D}%
q\right\rangle \right| ^{-1}\,,
\end{equation}
then the sufficiently intensive wave occurs unstable with respect to small
amplitude disturbance. The latter grows  and the wave inevitably breaks into
a chain of energy slots (solitons).

According to the instability conditions (22), if friction was absent then
sufficiently smooth spatial modulation always would unstable. Due to
friction, however, both long and short modulations always are stable (taking
into account that in reality\,  \,  \,$\,A<1\,$\,, because of
relations (9)). The instability starts from moderate modulation scales  \,$\,%
q_{c}\,$\,, and after exceeding at least minimum threshold amplitude
value, \,$\,A_{\min }\,\,$\,, as determined by (22),

\begin{equation}
\left| \left\langle q_{c}\,\widehat{D}q_{c}\right\rangle \right| \approx
\Gamma \,,\,   A_{\min }\approx \sqrt{\frac{\Gamma }{|\varkappa |}}
\end{equation}

5.6. SOLITONS.

Magnetic envelope soliton is a single spatially local wave packet stabilized
(protected from diffusional bleed) by non-linearity and described by Eq.5 or
Eq.11. Its existence is implied by the same instability (they must satisfy
the first of the inequalities (22)), but to created it one should use
suitably localized external pump, instead of a spreaded wave. These solitons
are called also ``bright solitons''.

The envelope of one-dimensional (flat) autonomous (at no pump at no
friction) bright soliton, moving in direction of some unit-length vector \,$\,%
\overline{n}\,\,$\,, is determined by the equations

\begin{equation}
\psi (\rho,t)=e^{-i\Omega t}\digamma (\xi )\,,\, \xi \equiv
\left\langle \overline{n}\,\rho -v_{g}t\right\rangle \,,\, \delta \frac{%
d^{2}\digamma }{d\xi ^{2}}-\varkappa \digamma ^{3}+\Omega \digamma =0\,,\,\
 \delta \equiv \left\langle \overline{n}\,\widehat{D}\overline{n}%
\right\rangle \,,\
\end{equation}
which directly follow from Eq.11. The solution to (24) is

\begin{equation}
\digamma (\xi )=A/\cosh (A\xi |\varkappa /2\delta |^{1/2})\,\,,\, \, \
\Omega =\varkappa \delta A^{2}/2\,,
\end{equation}
with the magnitude \,$\,A\,\,$\,  being free parameter.

Alternatively, so-called ``black'' (dark) envelope solitons can exist
representing ``holes'' (dips) in amplitude of spreaded (plane) wave. From
the point of view of above consideration, these objects formally correspond
to imaginary modulation wave vector, \,$\,q\rightarrow iq\,\,$\,. Thus for them the
first of the instability conditions (22) turns into opposite, but the second
remains valid. Hence, sufficiently intensive MW inevitably loses stability
and produces some soliton structure by either one or another way.

The envelope of black soliton satisfies the same equations (24), but with
non-zero boundary values at infinity, and has the form

\begin{equation}
\digamma (\xi )=A\tanh (A\xi |\varkappa /2\delta |^{1/2})\,\,,\, \, \
\Omega =\varkappa A^{2}
\end{equation}
Clearly, both the types of solitons are as much narrow (wide) as strong
(weak). The peculiarity of the black soliton is phase slip in its center by \,$\,%
\pi \,$\,. For a given direction \,$\,\overline{n}\,$\,, either bright (if \,$\,\,\delta
\varkappa <0\,$\, ) or black (if \,$\,\,\delta \varkappa >0\,$\, ) solitons exist only.
In reality, two-dimensional solitons are under use [3-9], but their
analytical investigation is much more hard task.

5.7. MAGNETIC CHAOS.

The comprehend reviews of experimental data on magnetic chaos  its
theoretical interpretation and numerical reproduction are presented in
[2,17,18,19]. The important conclusion from both theory and numerical
simulations is that even two MW modes (i.e. four variables: two amplitudes
and two phases) are sufficient to realize chaotic behavior.

In standard scenario, the mechanism of chaos is dependence of frequencies of
the modes on amplitudes  because of their nonlinear self-interaction and
mutual parametric interaction. At small amplitudes  they are coherently
(resonantly or parametrically) excited by external field and one by another.
At large amplitudes  the coherence destroys and dissipative damping prevails
which restores coherent interaction and returns to beginning of the cycle.
Under sufficiently strong pump, this cycle becomes unstable with respect to
infinitely small perturbation and thus chaotic.

Hence, the same property (non-isochronity) of magnetization oscillations at
the same degree of nonlinearity is responsible for both chaos in only
two-mode model and for regular soliton structures consisting of very many MW
modes (let us recollect that autonomous NSE dynamics is integrable and thus
can not produce chaos [20]). This fact demonstrates that principal origin of
magnetic chaos is nothing but energy transfer through magnetic system (from
external source to thermostat). In other words  this is dissipative chaos
characterized by phase volume contraction and dissipative strange (zero
Lebesgue measure) attractors [21], although (due to small friction)
possessing many properties of Hamiltonian chaos [21]. Then, it is not
surprising that just the power absorption (energy consumption by a
ferromagnet sample per unit time) mostly highlights magnetic chaos [2,17,18]
and can be used as control variable for its identification and
synchronization [19].

In principle, in presence of periodic perturbation accompanied by
dissipation even an individual spin (magnetic moment with fixed length and
thus two independent variables only) can undergo chaotic behavior [22-24].
Nevertheless  the two MW modes are too few to adequately imitate real
magnetic chaos since they are forced to incur roles of other modes.
Therefore a variety of many-mode models was suggested for numerical
investigation [2] which are able to reproduce (i) typically observed spikes
in the power time series  (ii) their intermittency, (iii) their fractal
properties and, moreover, (iv) characteristic frequencies of chaotic power
oscillations  usually in the interval from 0.1 MHz to 10 MHz. \

5.8. FRACTAL DIMENSION AND CONTROL OF CHAOS.

Physically, the peculiarity of chaos (in contrary to noise) is that very
many degrees of freedom are governed by a few independent variables only.
What anybody needs in when describing chaos is adequate choice of such the
relevant variables (which may differ from some particular modes). At least
the number of relevant variables  \,$\,d_{rel}\,\,$\,, can be determined if estimate
so-called fractal dimension, \,$\,d_{frac}\,$\,,\,$\,\,$\,of time series under
observation.

The quantity \,$\,d_{frac}\,\,$\,characterizes dimension of a manifold (attractor)
filled by trajectories of the relevant variables. It is obvious that \,$\,%
\,d_{frac}<d_{rel}\,\,$\,. At the same time, \,$\,\,d_{frac}\,>d_{rel}-1\,\,$\,, since
the opposite case would mean that one of variables is somehow dependent on
others. For example, if the attractor was periodic (limit) cycle whose
dimension \,$\,d_{frac}\,=1\,$\, it would be described by single variable (its
phase). For chaotic (strange) attractor, its dimensionality \,$\,d_{frac}\,\,$\, is
inevitably non-integer. This means that its intersection with (almost any)
one-dimensional line (in \,$\,\,d_{rel}\,$\,-dimensional embedding space) represents
so-called Cantor set. The latter is infinitely rarefied (nowhere dense) set
of uncountably many points [21]. Roughly speaking, if line contains \,$\,\aleph \,$\,
points then a Cantor set on it contains \,$\,\aleph ^{\delta }\,\,$\,points with \,$\,%
\,\delta <1\,$\,. Then \,$\,d_{frac}=\,$\, \,$\,d_{rel}-1+\delta \,$\,.

If \,$\,d_{frac}\,\,$\,is known then the number, \,$\,\,d_{rel}\,$\,, of variables which
are governing chaotic dynamics can be found as the integer number exceeding \,$\,%
d_{frac}\,\,$\,but most close to it. The cases when \,$\,d_{frac}\geq 3\,\,$\, (and thus
\,$\,d_{rel}\geq 4\,\,$\,) are called hyperchaos. For the same purpose of \,$\,\,d_{rel}\,$\,
determination, the correlation dimension can be used.


The correlation dimension [25], \,$\,d_{cor}\,\,$\,, characterizes statistics of
distances between points of the attractor taken at discrete time moments \,$\,%
t_{n}=t_{0}+n\tau \,$\,, with some reasonable time interval \,$\,\,\tau
\, \,$\,and \,$\,\,n=1..N\,$\,, at \,$\,N\rightarrow \infty \, \,$\,. Let \,$\,\,X_{d}(t)\,$\,, \,$\,%
d=1..d_{rel}\,$\,, be relevant variables under consideration. Then the set of
distancies

\[
R_{ij}=\{\sum_{d}[X_{d}(t_{i})-X_{d}(t_{j})]^{2}\}^{1/2}
\]
\,$\,\,$\,is investigated as characterized by the so-called correlation sum,

\[
\sigma (R)\equiv \frac{2}{N(N-1)}\sum_{1\leq i<j\leq N}\{R_{ij}<R\}\,,\,\
 \text{where\,  }\,   \{R_{ij}<R\}\equiv
\begin{array}{c}
1\,\text{  if\,  }\, R_{ij}<R \\
0\,\text{  if\,  }\, R_{ij}\geq R
\end{array}
\]

Due to fractal (scale-invariant) structure of Cantor sets  it can be
expected that at small distances and large number of points the power law
takes place:

\begin{equation}
\sigma (R)\rightarrow \Omega \left( \frac{R}{R_{\max }}\right)
^{d_{cor}}\, \,\text{at\,  \,  }\, N\rightarrow \infty \,\
,\,  \frac{R}{R_{\max }}\rightarrow 0\,,\, \, R_{\max }\equiv
\max_{ij}\,R_{ij}\,,
\end{equation}
where \,$\,\Omega \,$\, is some constant, and the right-hand limit of the
correlation sum presents definition of \,$\,\,d_{cor}\,$\,. Naturally, under rather
general assumptions \,$\,d_{cor}=\,$\, \,$\,d_{frac}\,\,$\,[25-28].

There are two important statements. First, \,$\,d_{cor}\, \,$\,is insensible to
smooth transformations of attractor variables including (not too long) time
delays. Therefore, equivalently one may analyze discrete sequence, \,$\,%
x(t_{0}+n\tau )\;\,$\,, of any available variable, \,$\,x(t)\;\,$\,(of course, well
connected to attractor), considering the subsequences \,$\,\{x(t_{0}+n\tau ),\,$\, \,$\,%
x(t_{0}+n\tau +\tau ),\,$\, \,$\,...,x(t_{0}+n\tau +d\tau )\}\,\,$\,, with \,$\,d\geq
d_{rel}\,\,$\,, quite like the attractor points above. Second, in principle,
estimate of \,$\,d_{cor}\, \,$\,is insensible to \,$\,d\,$\, (called embedding dimension)
if only \,$\,d\geq d_{rel}\,\,$\,. But it is sensible to noise, either external
parasitic one (errors of measurements  etc.) or ham noise produced by a real
system itself. Hence, a factual \,$\,d_{cor}\,$\,'s dependence on \,$\,d\,$\, can inform
about quality of data under analysis.

From the other hand, in general structure of chaos (strange attractor) may
be better characterized by a spectrum of fractal dimensions instead of a
single one [27,28]. Then different variables may give more or less different
correlation dimensions.


The essence of chaotic motion is its exponential instability, that is
exponential growth of response to arbitrarily small disturbance.
Nevertheless  this motion obey deterministic law. Hence, if its current
state is controlled with accuracy up to \,$\,n\,$\, binary digits then its future
can be somehow predicted for a time, \,$\,n\cdot t_{\inf }\,\,$\,, proportional to \,$\,%
n\,\,$\,. Then \,$\,\,h_{ch}=\ln 2/t_{\inf }\, \,$\,is called entropy of chaotic
attractor, while the sense of \,$\,\,t_{\inf }\,\,$\, is lifetime of information
bits. The latter approximately coincides with characteristic correlation
time of chaotic variables [21,33].


Let there are two identical chaotic generators initially delivered in the
same state to some extent of precision. To keep the same equality of states
in future and thus synchronize one generator by another, we should send from
one to another at least one bit of information per time \,$\,t_{\inf }\,\,$\,. A
representative chaotic variable carry just such the amount of information
and hence can be used for the synchronization [29,30]. The real example of
magnetic chaos synchronization was reported in [19].

However, to make this minimum necessary information to be also factually
sufficient, it should be chosen and applied in adequate way. Concretely, one
must take into account the topology of attractors  i.e. graph of transitions
between its Cantor subsets. For example [34], the discrete-time chaotic
evolution described by the tent map, \,$\,x(t+1)=\,$\, \,$\,1-|2x(t)-1|\,\,$\,, with \,$\,%
0<x<1 \,$\,, has \,$\,\,h_{ch}=\ln 2\,\,$\,, i.e. one bit of information per time step
is sufficient for synchronization. But this principal possibility turns into
reality if only the bit is chosen be \,$\,0\,$\, at \,$\,x<x_{0}\,$\, and \,$\,1\,$\, at \,$\,x>x_{0}\,$\,,
with certainly \,$\,x_{0}=1/2\,$\,. Any other rule (or other \,$\,x_{0}\,$\, ) either leads
to errors or requires additional information. Generally, determination of
adequate rule (termed generating partition of phase space) and corresponding
most meaningful information sequences (so-called symbolic dynamics) is very
non-trivial task [34,35], even if dynamic law of chaos is known, all the
more if it is under question. From this point of view, the results of [19]
seem extremely interesting.


In principle, an adequate rule allows to synchronize non-identical
attractors too if they have similar topologies and equal entropies. At more
simplified approaches to synchronization (but instead practically applied
ones  see [31,32] and references therein), rather small non-identity of
``master'' and ``slave'' chaotic systems can forbid it, even in spite of
quantitative excess of information.

Let \,$\,X(t)\,$\, and \,$\,Y(t)\,$\, are vector (\,$\,d_{rel}\,$\,-dimensional) variables of two
chaotic systems which obey the same dynamic equations but the second is
influenced by the first as follows:

\begin{equation}
dX/dt=F(X)\,\,,\,  dY/dt=F(Y)-g\cdot (Y-X)
\end{equation}
Here \,$\,g\,$\, is positive matrix, hence, it introduces additional damping. Let
the latter is so strong that suppresses exponential instability of the slave
system. Then it easy to see that after some time the only possible solution
for \,$\,Y(t)\,$\, will exactly reproduce \,$\,X(t)\,$\,, and thus one can say that the
slave system is ideally synchronized by the master system.

Perhaps  however, it would be more correct to name this copying chaos.
Indeed, the slave factually loses its autonomy (since at \,$\,X=0\,$\, it would
produce neither chaos nor any other motion instead tending to a stable
state), and behaves as passive repeater of external signal.

In more general and fine variant of such kind of synchronization,

\begin{equation}
dX/dt=F(X_{1}\,X)\,\,,\,  dY/dt=F(X_{1}\,Y)\,,
\end{equation}
where \,$\,X_{1}\,$\, is some (say first) of \,$\,d_{rel}\,$\, attractor variables  and
again function \,$\,F(X_{1}\,Y)\,$\, of two arguments is arranged so that first
equation produces chaos  while solution of the second falls into stable
point solution as \,$\,X_{1}=0\,$\,.


The copying of chaos is rather sensible to non-identity of the slave and
master systems  to adding external noise or any distortion of master signal
in transmission channel (see, for instance, [36] and references therein).
Relative error of the reproduction occurs be at least the same as relative
difference of master and slave parameters  plus noise to signal ratio and
plus relative distortions.

In practical applications  more reasonable approach may be to surely
recognize and reproduce some particular characteristics of chaotic signal
only, instead of its literal but erroneous copying. The example is mutual
phase synchronization of chaotic oscillators (for instance, famous Rossler
systems) which does not need in simultaneous amplitude synchronization and
therefore is possible for non-identical oscillators in presence of noise.
More general possibility is so-called event synchronization where events
mean definite well characterizable fragments of chaotic trajectory.

Then the natural step is an artificial creation of events in master system
which can serve for encoding information and then its decoding in similar
slave system. In particular, this may be switching between different
trajectories on the same attractor.


When artificially manipulating trajectory of a chaotic system, one needs in
a set (alphabet) of easy creatable and identifiable ``events''. Such the
possibility is ensured by unstable periodic orbits (UPO), i.e. periodic
trajectories  which always take place on chaotic attractors. Moreover, from
practical point of view one may treat a strange attractor merely as a
collections of periodic orbits with different length, from some minimum
period up to infinity. The instability of finite-length orbits means that
their measure (relative number of attractor points belonging them) is zero,
therefore, almost even insignificant deviation from short periodic orbit for
certain injects to very long one (chaotic).

But, remarkably, UPO's can be stabilized and thus practically installed into
master's chaotic trajectory by means of specially programmed feedback (see,
for example, [37]). Then similar feedback in slave system helps to
unambiguously recognize an UPO's installation although it looks quite as
typical fragment of transmitted signal. Such the discrete chaotic encryption
of information can well protect it from noise and signal distortions.


During recent decade many ideas of chaos application to secure communication
were suggested. One of schemes successfully realized in [31] is based on
introducing communication signal, \,$\,s(t)\,$\,, into Eqs.29 :

\begin{equation}
dX/dt=F(X_{1}+s(t),X)\,\,,\,  dY/dt=F(X_{1}+s(t),Y)\,
\end{equation}
Thus the master (in [31] it is chaotic Chua's generator) produces chaos
influenced by the signal. The factually transmitted information is the sum \,$\,%
\,s_{trans}(t)=\,$\, \,$\,X_{1}(t)+s(t)\,$\,. If the communication \,$\,s(t)\,$\, was absent \,$\,%
Y(t)\,$\, would be precise copy of \,$\,X(t)\,$\,. Therefore the communication can be
restored as \,$\, s(t)=s_{trans}(t)-Y_{1}(t)\,\,$\,.

Analogously, discrete chaos (chaotic maps) can be used. For simplest
example, let \,  \,$\,x(t+1)=\,$\, \,$\,F(x(t))\,$\, be some one-dimensional chaotic
map, and we introduce discrete-time information \,$\,s(t)\,$\, by means of
\,  \,$\,x(t+1)=\,$\, \,$\,F(x(t)+s(t))\,$\,. If \,$\,\,s_{trans}(t)\equiv \,$\, \,$\,x(t)+s(t)\,$\,
is sent, then in identical receiving system the information can be recovered
merely as \,$\,\,s(t)=\,$\, \,$\,\,s_{trans}(t)-\,$\, \,$\,F(s_{trans}(t-1))\,$\,. This
possibility was suggested in [38] for information encoding in chaotic
impulse communication.

Principally similar ideas were experimentally realized for communication
with optical chaos [39]. The peculiarity of the latter is an essential time
delay in a feedback part of optical (laser) chaotic generators.
Correspondingly, their dynamics undergo difference-differential nonlinear
equations which can produce chaos whose fractal dimensionality, \,$\,d_{frac}\,\,$\,%
, exceeds formal number of variables (number of equations).

Such the schemes (in which an information is either masked by chaos or
modulates it) possess all the potential defects of chaos copying. Besides
they do not allow for multiple-access chaotic communication (many senders)
in the same time-frequency domain. The matter is that chaotic system can not
recognize even its own signal if is mixed with a signal from other system.

Still chaotic extension of modern digital code-division multiple access
(CDMA) communication is under discussion. Although CDMA also uses chaotic
signals (pseudo-random coding sequences) but these are discrete exactly
predictable (periodic) signals only whose entropy is zero. The interesting
scheme of multiplexed chaotic communication based on analog chaotic signals
was suggested in [40,41]. It shows how the unrecognizability of mixed
chaotic messages can be overcame. The idea is that all the users
simultaneously take part in creating chaos which thus becomes common for all
the network and therefore recognizable by any participating chaotic
generator. At present form, however, this scheme needs in temporal
separation of users and other limitations [40,41].

{\it REFERENCES}

1. A.I.Akhiezer, V.G.Baryakhtar and S.V.Peletminski. Spin waves. Moscow,
Nauka Publ., 1967.

2. Nonlinear phenomena and chaos in magnetic materials. Editor Ph.E.Wigen.
World Sci. Publ., 1994.

3. A.N.Slavin, B.A.Kalinikos and N.G.Kovshikov. In Ref. 2, p. 209.

4. B.A.Kalinikos  N.G.Kovshikov and C.E.Patton. Phys.Rev.Lett. 80 (1998)
4301.

5. Hua Xia, P.Kabos  Hong Yan Zhang, P.A.Kolodin and C.E.Patton.
Phys.Rev.Lett. 81 (1998) 449.

6. Hua Xia, P.Kabos  R.A.Staudinger and C.E.Patton. Phys.Rev. B58 (1998)
2708.

7. A.N.Slavin, Yu.S.Kivshar, E.A.Ostrovskaya and H.Benner. Phys.Rev.Lett. 82
(1999) 2583.

8. B.A.Kalinikos  M.M.Scott and C.E.Patton. Phys.Rev.Lett. 84 (2000) 4697.

9. O.Buttner, M.Bauer, S.O.Demokritov, at al. Phys.Rev. B61 (2000) 11576.

10. P.A.Kolodin, P.Kabos and C.E.Patton. Phys.Rev.Lett. 80 (1998) 1976.

11. G.A.Melkov, A.A.Serga, V.S.Tiberkevich and A.N.Oliynyk. Phys.Rev.Lett.
84 (2000) 3438.

12. G.A.Melkov, Yu.V.Kobljanskyj, A.A.Serga and V.S.Tiberkevich.
Phys.Rev.Lett. 86 (2001) 4918.

13. B.C.Choi, M.Belov, V.K.Hiebert, at al. Phys.Rev.Lett. 86 (2001) 728.

14. M.Bauer, O.Buttner, S.O.Demokritov and B.Hillebrands. Phys.Rev.Lett. 81
(1998) 3769.

15. V.T.Synogach, Yu.K.Fetisov, C.Mathieu and C.E.Patton. Phys.Rev.Lett. 85
(2000) 2184.

16. R.A.Kraenkel, M.A.Manna and V.Merle. Phys.Rev. B61 (2000) 976.

17. S.M.Rezende and F.M.de Aguiar. Proc. IEEE, 78 (1990) 893.

18. J.Beeker, F.Rodelsperger,Th.Weyrauch, H.Benner, W.Just and A.Cenys.
Phys.Rev. E59 (1999) 1622.

19. D.W.Peterman, M.Ye and P.E.Wigen. Phys.Rev.Lett. 74 (1995) 1740.

20. G.B.Whitham. Linear and nonlinear waves. Wiley Intersci. Publ., 1974.

21. A.J.Lichtenberg and M.A.Lieberman. Regular and stochastic motion.
Springer-Verlag, 1988.

22. L.F.Alvarez, O.Pla and O.Chubykalo. Phys.Rev. B61 (2000) 11613.

23. G.Bertotti, C.Serpico and I.D.Maeyrgoyz. Phys.Rev.Lett., 86 (2001) 724.

24. G.Bertotti, I.D.Maeyrgoyz and C.Serpico. Phys.Rev.Lett., 87 (2001) 7203.

25. P.Grassberger and I.Procaccia. Physica, D9 (1983) 189.

26. H.G.E.Hentschel and I.Procaccia. Physica, D8 (1983) 435.

27. H.Yamazaki. Fractal properties of magnetic crystal. In Ref. 2, p. 191.

28. T.S.Akhromeeva, S.P.Kurdyumov, G.G.Malinetskii and A.A.Samarskii. Chaos
and dissipative structures in reaction-diffusion systems. Nauka Publ.,
Moscow, 1992.

29. L.M.Pecora and T.L.Carrol. Phys.Rev.Lett., 64 (1990) 821.

30. E.Ott, C.Grebogi and J.A.Yorke. Phys.Rev.Lett., 64 (1990) 1196.

31. A.S.Dmitriev, A.I.Panas and S.O.Starkov. Int.J.Bif.\&Chaos  7 (1997)
2511.

32. T.Yang and L.O.Chua. Int.J.Bif.\&Chaos  7 (1997) 2789.

33. B.Chirikov. Linear and nonlinear chaos. Chao-dyn/9705003.

34. E.M.Bolt, Th.Stanford, Ying-Cheng Lai and K.Zyczkowski. Phys.Rev.Lett.,
85 (2000) 3524.

35. R.L.Davidchack, Ying-Cheng Lai, E.M.Bolt and M.Dhamala. Phys.Rev., E61
(2000) 1353.

36. N.F.Rulkov and L.S.Tsimring,\, arXiv: chao-dyn/9705019.

37. K.Pyragas. Phys.Rev.Lett., 86 (2001) 2265.

38. N.F.Rulkov, M.M.Sushchik, L.S.Tsimring, et al., %
arXiv: chao-dyn/9908015.

39. V.S.Udaltsov, J.-P. Goedgebuer, L.Larger and W.T.Rhodes. Phys.Rev.Lett.,
86 (2001) 1892.

40. K.Yoshimura. Phys.Rev., E60 (1999) 1648.

41. Sh.Sundar and A.A.Minai. Phys.Rev.Lett., 85 (2000) 5456.



\,\,\,

\,\,\, 

\section*{Conclusion}

We hope that at least some parts %
of the aforesaid material can be useful %
supplement to existing literature %
on magnetic waves. Anyway, the presented approach, - %
based on first principles only, - %
well helps to understand and interpret results of %
numerical simulation of linear and non-linear magnetostatic %
waves and magnetic chaos. This will be subject %
of continuation of this manuscript.

\,\,\,

\,\,\,

--------------------------------------------

\,\,\,


\end{document}